\documentclass[12pt]{article}
\usepackage{a4}
\usepackage{array}
\usepackage{calc}
\newlength{\adressabstand}
\usepackage[PostScript=dvips]{diagrams}
\diagramstyle[Postscript=dvips]
\newarrow{auf}|---{->}
\newarrow{nach}----{->}
\newarrow{ident}33333
\newarrow{aufsurj}|---{->>}
\newarrow{nachsurj}----{->>}
\newarrow{einbett}C---{->}
\newarrow{strichel}{}{dash}{}{dash}{->}
\newarrow{impliz}===={=>}
\newlength{\CDhoehe}                  
\newlength{\CDgap}                    
\usepackage{ifthen,array}
\newcommand{\ams}{\usepackage{amsfonts,amssymb,amsmath}}

\ams
\allowdisplaybreaks[3]
\newlength{\textwidthorig}
\newlength{\oddsidemarginorig}
\newlength{\textheightorig}
\newlength{\topmarginorig}
\def\seitenlaengenabsolut#1 #2 #3 #4 {\setlength{\textwidth}{#1}
                                      \setlength{\oddsidemargin}{#2}
                                      \setlength{\textheight}{#3}
                                      \setlength{\topmargin}{#4}}
\def\seitenlaengenrelzustandard#1 #2 #3 #4 {\setlength{\textwidth}{\textwidthorig+#1}
                                            \setlength{\oddsidemargin}{\oddsidemarginorig+#2}
                                            \setlength{\textheight}{\textheightorig+#3}
                                            \setlength{\topmargin}{\topmarginorig+#4}}
\def\seitenlaengenrelzuvorher#1 #2 #3 #4 {\addtolength{\textwidth}{#1}
                                          \addtolength{\oddsidemargin}{#2}
                                          \addtolength{\textheight}{#3}
                                          \addtolength{\topmargin}{#4}}
\newcommand{\standardseite}{\seitenlaengenrelzuvorher2.2cm -0.8cm 1.8cm -1.5cm }   
\standardseite
\newcommand{\leerezeile}{\vspace{2ex}}
\newlength{\laengespatium}

\newcommand{\nach}{\longrightarrow}      
\newcommand{\txtnach}[1]{\xrightarrow{#1}}
\newcommand{\auf}{\longmapsto}           
\newcommand{\txtauf}[1]{\auf}            
\newcommand{\impliz}{\Longrightarrow}    
\newcommand{\aequ}{\Longleftrightarrow}  
 
\newcommand{\invimpliz}{\Longleftarrow}  
\newcommand{\gegen}{\rightarrow}         
\newcommand{\iso}{\cong}                 
\newcommand{\ident}{\equiv}              
\newcommand{\teilmenge}{\subseteq}       
\newcommand{\obermenge}{\supseteq}       
\newcommand{\echteteilmenge}{\subset}    
\newcommand{\echteobermenge}{\supset}    
\newcommand{\aeqrel}{\sim}               
\newcommand{\nichtin}{\not\in}

\newcommand{\fueralle}{\hspace{1.7em}\forall}
\newcommand{\leeremenge}{\varnothing}    
\newcommand{\kreuz}{\times}              

\newcommand{\einschr}[1]{{}\arrowvert_{#1}}      
\newcommand{\dirsum}{\oplus}           
\newcommand{\dirprod}{\operatorname*{{\text{\Large$\boldsymbol{\kreuz}$}}}}
\newcommand{\kp}{\odot}                  
\newcommand{\betraganpass}[1]%
           {\left| #1 \right|}           
\newcommand{\bigbetrag}[1]%
           {\bigl|{#1}\bigr|}            
\newcommand{\betrag}[1]%
           {|{#1}|}                      
\newcommand{\betragnichtanpass}[1]%
           {\mid #1 \mid}                
\newcommand{\norm}[1]%
           {{}{\parallel}#1{\parallel}{}}      
\newcommand{\erww}[1]%
           {\langle #1 \rangle}          
\newcommand{\skalprod}[2]%
           {\langle #1,#2 \rangle}       
\newcommand{\quer}{\overline}            
\newcommand{\dach}{\widehat}             
\newcommand{\inv}[1]{\frac{1}{#1}}       
\newcommand{\einhalb}{\inv{2}}           
\newcommand{\re}{\text{Re }}                           
\newcommand{\ima}{\text{Im }}                          
\newcommand{\im}{\text{im\;}}                          
\newcommand{\pr}{{\text{pr}}}                          
\newcommand{\tr}{\text{tr}}                           
\newcommand{\ido}{\text{id}}                           
\newcommand{\inter}{\text{int}\:}                      
\newcommand{\Ad}{{\text{Ad}}}                          
\newcommand{\elanz}{\#}                                
\newcommand{\del}{\partial}                            
\newcommand{\Hom}{\text{Hom}}                          
\newcommand{\Maps}{\text{Maps}}                        
\newcommand{\dd}{\text{d}}                             
\newcommand{\e}{\text{e}}                              
\newcommand{\I}{\text{i}}                              
\newcommand{\J}{\text{j}}                              
\newcommand{\KK}{\text{k}}                             
\newcommand{\field}[1]{\mathbb{#1}}                    
\renewcommand{\H}{{\field{H}}}                         
\newcommand{\C}{{\field{C}}}                           
\newcommand{\N}{{\field{N}}}                           
\newcommand{\R}{{\field{R}}}                           
\newcommand{\Z}{{\field{Z}}}                           
\newcommand{\rnkl}[2]{\raisebox{-0.4ex}{$#1$}%
\raisebox{-0.12ex}{{\large$\setminus$}}\,#2}   
\newcommand{\agb}{{\overline{{\cal A}/{\cal G}}}}      
\newcommand{\agbfact}[1][]{\text{$\agb/\!\aeqrel$}}    
\newcommand{\ag}{{\cal A}/{\cal G}}                    
\newcommand{\Ab}{{\overline{{\cal A}}}}                
\newcommand{\A}{{\cal A}}                              
\newcommand{\Gb}{{\overline{{\cal G}}}}                
\newcommand{\Gbconst}{{\Gb_{\text{const}}}}
\newcommand{\AbGb}{{\Ab/\Gb}}                          
\newcommand{\gen}{{\text{gen}}}
\newcommand{\Abgen}{{\Ab_\gen}}                        
\newcommand{\AbGbgen}{{\Abgen/\Gb}}                    
\newcommand{\AbGbeq}[1]{\Abeq{#1}/\Gb}                 
\newcommand{\qa}{{\quer{A}}}                           
\newcommand{\qg}{{\quer{g}}}                           
\newcommand{\hg}{{\cal HG}}                            

\newcommand{\holgr}{{\mathbf H}}                       
\newcommand{\bz}{{\mathbf B}}                          
\newcommand{\bzz}{{\bz_Z}}                             
\newcommand{\nklna}{\rnkl{\bz(\qa)}{N(\bz(\qa))}}      
\newcommand{\nklnza}{\rnkl{Z(\holgr_\qa)}{N(Z(\holgr_\qa))}}
\newcommand{\nklagen}{\rnkl{\bzz}{\Gb}}
\newcommand{\nklagenab}{\rnkl{\Gbconst}{\Gb}}
\newcommand{\nknonabel}{\rnkl{Z}{\LG}}
\newcommand{\GR}{\Gamma}                               
\newcommand{\Ver}{\mathbf{V}}                          
\newcommand{\Edg}{\mathbf{E}}                          
\newcommand{\gross}[1]{{\boldsymbol #1}}               
\newcommand{\ga}{\gross{\alpha}}                       
\newcommand{\gb}{\gross{\beta}}                        
\newcommand{\gc}{\gross{\gamma}}                       
\newcommand{\pf}[1]{{\cal P}_{#1}}                     
\newcommand{\Pf}{{\cal P}}                             
\newcommand{\KG}[1]{\Pf_{#1}}                          
\newcommand{\hyph}{\upsilon}                           
\newcommand{\Hyph}{Y}                                  
\newcommand{\Haar}{{\text{Haar}}}                      
\newcommand{\LG}{{\mathbf{G}}}                         
\newcommand{\LH}{{\mathbf{H}}}                         
\newcommand{\LK}{{\mathbf{K}}}                         
\newcommand{\aeqrelzush}[1][]{\sim}                    
\newcommand{\Abeq}[1]{\Ab_{=#1}}                       
\newcommand{\typ}{\text{Typ}}                          
\newcommand{\qS}{{\quer S}}                            
\newcommand{\qz}{{\quer z}}                            
\newcommand{\redisoeo}{\Psi_0}                         
\newcommand{\T}{{\cal T}}
\newcommand{\nklza}[1][]{\ifthenelse{\equal{#1}{}}     
                                    {\rnkl{Z(\holgr_\qa)}{\LG}}        
                                   {\rnkl{Z(\holgr_{#1})}{\LG}}}       
\newcommand{\nkla}[1][]{\ifthenelse{\equal{#1}{}}      
                                    {\rnkl{\bz(\qa)}{\Gb}}        
                                    {\rnkl{\bz(#1)}{\Gb}}}       
\newcommand{\ueberd}{{\cal U}}                         

\newcommand{\vg}{\vec g}
\newcommand{\YM}{{\text{YM}}}                          

\newcommand{\ymwirk}[1][]{\ifthenelse{\equal{#1}{}}{S_{\YM}}{S_{\YM,#1}}}

\newcommand{\DD}{{\cal D}}                             
\newcommand{\fpdet}{\Delta}                            
\newcommand{\loktriv}{\chi}                            
\newcommand{\Iota}{I}
\newcommand{\kmin}{k_{\min}}
\newcommand{\vv}{\vec v}
\newcommand{\ueberdv}{{\cal V}}
\newcommand{\punkt}{\ast}
\newcommand{\ueberdabgen}{{\cal U}}                    
\newcommand{\svp}{s_\varphi}
\newcommand{\svpf}{s_{[\varphi]}}

\newcommand{\homo}{{\text{h}}}
\newcommand{\pt}{{\text{pt}}}
\newcommand{\Stab}[2][G]{\text{Stab}(#2)} 
\newcommand{\isonorm}{{\psi}}
\newcommand{\isonorma}{{\chi}}
\newcommand{\isonormb}{{\varphi}}
\newcommand{\smatrix}[1]{\bigl( \begin{smallmatrix} #1 \end{smallmatrix} \bigr)}
\newcommand{\EM}{{\mathbf{1}}}
\newcommand{\bmat}{\begin{pmatrix}}
\newcommand{\emat}{\end{pmatrix}}
\newcommand{\almgen}{{\text{alm gen}}}

\newcommand{\irred}{{\text{irr}}}
\newcommand{\Homeo}{{\text{Homeo}}}                      
\newcommand{\uebergabb}{\rho}              
\newcommand{\codim}{\text{codim}}
\newcommand{\klammerunten}[2]{\underbrace{#1}_{\textstyle#2}}
\newcommand{\genueberschrift}{{\mathrm{gen}}}
\newcommand{\ListNullAbstaende}{\setlength{\topsep}{0pt}%
                                \setlength{\parskip}{0pt}%
                                \setlength{\partopsep}{0pt}%
                                \setlength{\itemsep}{0pt}%
                                \setlength{\parsep}{0pt}}
\newcommand{\ListNurAnstrichAbstand}{\setlength{\topsep}{0pt}%
                                     \setlength{\parskip}{0pt}%
                                     \setlength{\partopsep}{0pt}%
                                     \setlength{\parsep}{0pt}}
\newenvironment{StandardListe}[2]%
               {\begin{list}%
                      {#1}%
                      {\settowidth{\leftmargin}{M#1}%
                       \settowidth{\labelwidth}{#1}%
                       \settowidth{\labelsep}{M}%
                       #2%
                      }%
                }%
               {\end{list}}%
\newenvironment{EinfachListe}[1]%
               {\begin{StandardListe}{#1}{\ListNullAbstaende}}%
               {\end{StandardListe}}%
               {\begin{StandardListe}{#1}{\ListNurAnstrichAbstand}}%
               {\end{StandardListe}}%
\newcommand{\labelsatz}[1]{#1}
\newcounter{listennr}                     
\newlength{\hilfslaenge}
\newlength{\stdlabellaenge}
\newlength{\maximum}
\newcommand{\stdlabel}{}
\newcommand{\Maximum}{}
\newcommand{\iitem}[1][]{\ifthenelse{\equal{#1}{}}%
                           {\item \setlength{\hilfslaenge}{\stdlabellaenge}}%
                           {\item[\labelsatz{#1}\hfill]%
                            \settowidth{\hilfslaenge}{\labelsatz{#1}}}%
                         \ifthenelse{\lengthtest{\maximum < \hilfslaenge}}%
                           {\setlength{\maximum}{\hilfslaenge}%
                            \ifthenelse{\equal{#1}{}}%
                               {\renewcommand{\Maximum}{\stdlabel}}%
                               {\renewcommand{\Maximum}{#1}}}%
                           {}%
                      }      
\makeatletter
\newenvironment{AutoLabelLaengenListe}[2][]%
               {\begin{list}%
                      {\labelsatz{#1}\hfill}%
                      {\stepcounter{listennr}%
                       \settowidth{\leftmargin}{M\labelsatz{\ref{listnr\arabic{listennr}}}}%
                       \settowidth{\labelwidth}{\labelsatz{\ref{listnr\arabic{listennr}}}}%
                       \settowidth{\labelsep}{M}%
                       \settowidth{\stdlabellaenge}{\labelsatz{#1}}%
                       \renewcommand{\stdlabel}{#1}%
                       #2%
                       \renewcommand{\Maximum}{}%
                      }%
                }%
               {\renewcommand{\@currentlabel}{\Maximum}%
                \label{listnr\arabic{listennr}}%
                \end{list}%
                }%
\makeatother
\newenvironment{StandardEinrueckung}[2]%
               {\begin{list}%
                      {#1}%
                      {\settowidth{\leftmargin}{M#1}%
                       \settowidth{\labelwidth}{#1}%
                       \settowidth{\labelsep}{M}%
                       #2%
                      }%
                \item}%
               {\end{list}}%
\newenvironment{Einrueckungpur}[1]%
               {\begin{StandardEinrueckung}{#1}{\ListNullAbstaende}}%
               {\end{StandardEinrueckung}}%
\newenvironment{Einrueckung}[1]%
               {\begin{StandardEinrueckung}{#1}{\setlength{\parsep}{0pt}}}%
               {\end{StandardEinrueckung}}%
\newcommand{\EineZeileGleichung}[2][0.0ex]
           {
            
            \vspace{#1} 
            \noindent
            \hspace*{\fill}
            $\displaystyle{#2}$
            \hspace*{\fill}

            \vspace{#1} 
            
           }
\newcommand{\EineZeileGleichungqed}[2][0.0ex]
           {
            
            \vspace{#1}%
            \noindent%
            \phantom{\qed}%
            \hspace*{\fill}%
            $\displaystyle{#2}$%
            \qed%

            \vspace{#1} 
            
           }
\makeatletter
\newcommand{\EineNumZeileGleichung}[2][0.5ex]
           {
            
            \vspace{#1} 
            \noindent
            \stepcounter{equation}
            \renewcommand{\@currentlabel}{\arabic{equation}}%
            \phantom{(\arabic{equation})}\hspace*{\fill}
            $\displaystyle{#2}$
            \hspace*{\fill}
            (\arabic{equation})

            \vspace{#1} 
            
           }
\makeatother
\makeatletter
\newcommand{\EineErwNumZeileGleichung}[2][0.5ex]
           {
            
            \vspace{#1} 
            \noindent
            \stepcounter{equation}
            \renewcommand{\@currentlabel}{\arabic{equation}}%
            \phantom{(\arabic{equation})}\hspace*{\fill}
            #2 
            \hspace*{\fill}
            (\arabic{equation})

            \vspace{#1} 
            
           }
\makeatother
\newcommand{\breitrel}[1]{\hspace*{\tabcolsep} #1 \hspace*{\tabcolsep}}
\newlength{\abstaug}              
\newenvironment{AllgUnnumGleichung}[2][1.0ex]%
               {
  
                \setlength{\abstaug}{#1}
                \vspace{\abstaug}
                \hspace*{\fill}
                $\begin{array}[t]{#2}
                }%
               {\end{array}$
                \hspace*{\fill}
  
                \vspace{\abstaug}

                }%
\newenvironment{AllgNumGleichung}[2][0.0ex]%
               {
  
                \setlength{\abstaug}{#1}
                \vspace{\abstaug}
                $\begin{tabular*}{\textwidth}[t]{#2}
                }%
               {\end{tabular*}$

                \vspace{\abstaug}

               }%
\newenvironment{StandardUnnumGleichungKlein}[1][0ex]
               {\renewcommand{\s}{\\[#1] }%
                \begin{AllgUnnumGleichung}{rcl}}%
               {\end{AllgUnnumGleichung}}%
\newcommand{\s}{\\[0ex] }             
\newenvironment{StandardUnnumGleichung}[1][0ex]
               {\renewcommand{\s}{\\[#1] }%
                \begin{AllgUnnumGleichung}{>{\displaystyle}rc>{\displaystyle}l}}%
               {\end{AllgUnnumGleichung}}%
\newenvironment{XrelYZNumGleichung}[1][0ex]%
               {\renewcommand{\s}{\\[#1] }%
                \begin{AllgNumGleichung}{rcll}}%
               {\end{AllgNumGleichung}}%
\newcommand{\erl}[1]{\hfill\mbox{\hspace*{1.5em}\small (#1)}}

\newcommand{\erllang}[2][0.5\textwidth]%
              {\hfill\hspace*{1.5em}%
               \begin{minipage}[t]{#1}{\small%
                          \begin{list}{(}{\ListNullAbstaende%
                                          \settowidth{\leftmargin}{(}%
                                          \settowidth{\labelwidth}{(}%
                                          \settowidth{\labelsep}{}%
                                         }%
                          \item#2)%
                          \end{list}}%
               \end{minipage}\\[-0.9ex]
              }%
\newcommand{\DefBemUmgeb}[1]%
           {\newenvironment{#1}[1][]%
                           {\begin{Einrueckung}{{\bf #1}}%
                            \ifx##1\empty\else{{\bf ##1}
                            
                                                        }\fi%
                            }%
                           {\end{Einrueckung}}}
\newcommand{\DefSBemUmgeb}[2]%
           {\newenvironment{#1}[1][]%
                           {\begin{Einrueckung}{{\bf #2}}%
                            \ifx##1\empty\else{{\bf ##1}
                            
                                                        }\fi%
                            }%
                           {\end{Einrueckung}}}
\makeatletter
\newcommand{\DefBspUmgeb}[3]%
           {\newcounter{#2}[#3]%
            \newenvironment{#1}[1][]%
                           {\stepcounter{#2}%
                            \renewcommand{\ZaehlerMarke}{\arabic{#2}}%
                            \renewcommand{\Einzugsname}{{\bf #1 \ZaehlerMarke}}%
                            \begin{Einrueckung}{\Einzugsname}
                            \ifx##1\empty\else{{\bf ##1}\\}\fi%
                            \renewcommand{\@currentlabel}{\ZaehlerMarke}%
                            }%
                           {\end{Einrueckung}}}
\makeatother
\newcommand{\ZaehlerbisEbene}{section}
\newcommand{\Ebenea}{section}
\newcommand{\Ebeneb}{subsection}

\newcommand{\Abschnittnummer}{%
            \ifx\ZaehlerbisEbene\Ebenea{\arabic{section}}%
             \else{%
              \ifx\ZaehlerbisEbene\Ebeneb{\arabic{section}.\arabic{subsection}}%
               \else{\arabic{section}.\arabic{subsection}.\arabic{subsubsection}}%
              \fi}%
            \fi}     
\newcommand{\Abschnittnummerpunkt}{\Abschnittnummer.}     
\newcommand{\Einzugsname}{}
\newcommand{\ZaehlerMarke}{}
\makeatletter
\newcommand{\DefThmUmgeb}[3]%
           {\newcounter{#1}[#3]%
            \newenvironment{#1}[1][]%
                           {\stepcounter{#2}%
                            \setcounter{#1}{\value{#2}}%
                            \renewcommand{\ZaehlerMarke}{\Abschnittnummerpunkt\arabic{#1}}%
                            \renewcommand{\Einzugsname}{{\bf #1 \ZaehlerMarke}}%
                            \begin{Einrueckung}{\Einzugsname}
                            \ifx##1\empty\else{{\bf ##1}
                            
                                                        }\fi%
                            \renewcommand{\@currentlabel}{\ZaehlerMarke}%
                            }%
                           {\end{Einrueckung}}}
\makeatother
\makeatletter
\newcommand{\DefSThmUmgeb}[4]%
           {\newcounter{#1}[#3]%
            \newenvironment{#1}[1][]%
                           {\stepcounter{#2}%
                            \setcounter{#1}{\value{#2}}%
                            \renewcommand{\ZaehlerMarke}{\Abschnittnummerpunkt\arabic{#1}}%
                            \renewcommand{\Einzugsname}{{\bf #4 \ZaehlerMarke}}
                            \begin{Einrueckung}{\Einzugsname}
                            \ifx##1\empty\else{{\bf ##1}

                                                        }\fi%
                            \renewcommand{\@currentlabel}{\ZaehlerMarke}%
                            }%
                           {\end{Einrueckung}}}
\makeatother
\makeatletter
\newcommand{\DefUnterNumThmUmgeb}[5]%
           {\newcounter{#1}[#3]%
            \newcounter{#4}%
            \newenvironment{#1}[1][]%
                           {\ifx##1\empty\else{\stepcounter{#2}\setcounter{#4}{0}}\fi%
                            \stepcounter{#4}%
                            \setcounter{#1}{\value{#2}}%
                            \renewcommand{\ZaehlerMarke}{\Abschnittnummerpunkt\arabic{#1}\alph{#4}}%
                            \renewcommand{\Einzugsname}{{\bf #5 \ZaehlerMarke}}
                            \begin{Einrueckung}{\Einzugsname}
                            \renewcommand{\@currentlabel}{\ZaehlerMarke}%
                            }%
                           {\end{Einrueckung}}}
\makeatother
\newenvironment{Beweis}[1][]%
               {\begin{Einrueckung}{{\bf Beweis}}%
                \ifx#1\empty\else{{\bf #1}

                                            }\fi%
                }%
               {\end{Einrueckung}%
                }%
\newenvironment{Proof}[1][]%
               {\begin{Einrueckung}{{\bf Proof}}%
                \ifx#1\empty\else{{\bf #1}

                                            }\fi%
                }%
               {\end{Einrueckung}%
                }%
               {\begin{Einrueckung}{{\bf \glqq Beweis\grqq}}%
                \ifx#1\empty\else{{\bf #1}
                
                                            }\fi%
                }%
               {\end{Einrueckung}%
                }%
               {\begin{Einrueckung}{{\bf Begr"undung}}%
                \ifx#1\empty\else{{\bf #1}
                
                                            }\fi%
                }%
               {\end{Einrueckung}%
                }%
\newenvironment{Hinrichtung}%
               {\begin{Einrueckungpur}{$\impliz$}}%
               {\end{Einrueckungpur}}%
\newenvironment{Rueckrichtung}%
               {\begin{Einrueckungpur}{$\invimpliz$}}%
               {\end{Einrueckungpur}}%
               {\begin{Einrueckungpur}{\glqq$\teilmenge$\grqq}}%
               {\end{Einrueckungpur}}%
               {\begin{Einrueckungpur}{\glqq$\obermenge$\grqq}}%
               {\end{Einrueckungpur}}%
               {\begin{Einrueckungpur}{"$\teilmenge$"}}%
               {\end{Einrueckungpur}}%
               {\begin{Einrueckungpur}{"$\obermenge$"}}%
               {\end{Einrueckungpur}}%
\newcommand{\qed}{\nopagebreak\hspace*{2em}\hspace*{\fill}{\bf qed}}
\newcommand{\ARabic}{\arabic}
\newcommand{\Nummerntypa}{\arabic}   
\newcommand{\Nummerntypb}{\alph}
\newcommand{\Nummerntypc}{\roman}
\newcommand{\Nummerntypd}{\Alph}

\newcommand{\Nra}{\Nummerntypa{Nummera}}            
\newcommand{\Nrb}{\Nummerntypb{Nummerb}}            
\newcommand{\Nrc}{\Nummerntypc{Nummerc}}                
\newcommand{\Nrd}{\Nummerntypd{Nummerd}}                
\newcommand{\ZeichenzuNrTyp}[1]%
           {\ifx#1\ARabic {.}\else{)}%
                  \fi}                              %
\newcommand{\NrZeicha}{\ZeichenzuNrTyp{\Nummerntypa}}
\newcommand{\NrZeichb}{\ZeichenzuNrTyp{\Nummerntypb}}
\newcommand{\NrZeichc}{\ZeichenzuNrTyp{\Nummerntypc}}
\newcommand{\NrZeichd}{\ZeichenzuNrTyp{\Nummerntypd}}
\newcommand{\ListMarkea}%
           {\Nra\NrZeicha}
\newcommand{\ListMarkeb}%
           {\Nra\NrZeicha\Nrb\NrZeichb}
\newcommand{\ListMarkec}%
           {\Nra\NrZeicha\Nrb\NrZeichb\Nrc\NrZeichc}
\newcommand{\ListMarked}%
           {\Nra\NrZeicha\Nrb\NrZeichb\Nrc\NrZeichc\Nrd\NrZeichd}
\newcommand{\Anfangszeichen}{}
\newcommand{\Anfangspunkt}{}
\newcounter{Schachtelebene}
\newcounter{Hilfszaehler}
\newcommand{\Hilfsbefehl}{}
\newcommand{\Schachtelebene}{\alph{Schachtelebene}}
\makeatletter
\newenvironment{AllgNumerierteListe}[2][]%
               {\addtocounter{Schachtelebene}{1}%
		\setcounter{Hilfszaehler}{#2}%
                \renewcommand{\Anfangszeichen}%
                             {\renewcommand{\Hilfsbefehl}{\csname Nummerntyp\Schachtelebene \endcsname}%
                              \Hilfsbefehl{Hilfszaehler}}%
                \renewcommand{\Anfangspunkt}%
                             {\csname NrZeich\Schachtelebene \endcsname}%
                \begin{list}%
                      {\stepcounter{Nummer\Schachtelebene}%
                       \csname Nr\Schachtelebene \endcsname
                       \csname NrZeich\Schachtelebene \endcsname
                       }%
                      {\settowidth{\leftmargin}{M\Anfangszeichen\Anfangspunkt}%
                       \settowidth{\labelwidth}{\Anfangszeichen\Anfangspunkt}%
                       \settowidth{\labelsep}{M}%
                       \setlength{\topsep}{0pt}%
                       \setlength{\parskip}{0pt}%
                       \setlength{\partopsep}{0pt}%
                       \setlength{\itemsep}{0pt}%
                       \setlength{\parsep}{0pt}%
                      }%
                \renewcommand{\@currentlabel}{\csname ListMarke\Schachtelebene \endcsname}%
                }%
               {\ifthenelse{\equal{}{}}{\setcounter{Nummer\Schachtelebene}{0}}{}
                \addtocounter{Schachtelebene}{-1}%
                \end{list}}
\makeatother
\newenvironment{NumerierteListe}[1]%
               {\begin{AllgNumerierteListe}{#1}}
               {\end{AllgNumerierteListe}}
\newenvironment{WeiterNumerierteListe}[1]%
               {\begin{AllgNumerierteListe}[Weiter]{#1}}
               {\end{AllgNumerierteListe}}

\newcommand{\UnnumAnfangszeichen}{}
\newcounter{UnnumSchachtelebene}
\newcommand{\UnnumSchachtelebene}{\alph{UnnumSchachtelebene}}
\makeatletter
\newenvironment{UnnumerierteListe}%
               {\addtocounter{UnnumSchachtelebene}{1}%
                \renewcommand{\UnnumAnfangszeichen}%
                             {\csname UnnumZeich\UnnumSchachtelebene \endcsname}%
                \begin{list}%
                      {\UnnumAnfangszeichen}%
                      {\settowidth{\leftmargin}{M\UnnumAnfangszeichen}%
                       \settowidth{\labelwidth}{\UnnumAnfangszeichen}%
                       \settowidth{\labelsep}{M}%
                       \setlength{\topsep}{0pt}%
                       \setlength{\parskip}{0pt}%
                       \setlength{\partopsep}{0pt}%
                       \setlength{\itemsep}{0pt}%
                       \setlength{\parsep}{0pt}%
                      }%
                }%
               {\addtocounter{UnnumSchachtelebene}{-1}%
                \end{list}}
\makeatother
\newlength{\fktdefhilfslaenge}
\newcommand{\ohnefktdef}[4]%
           {\hspace*{\fill}
            $\begin{array}[t]{ccc}%
            #1 & \nach & #2 \\
            #3 & \auf  & #4
            \end{array}$
            \hspace*{\fill}}
\newcommand{\fktdef}[5]%
           {\hspace*{\fill}
            $\begin{array}[t]{cccc}%
            #1: & #2 & \nach & #3 \\    
                & #4 & \auf  & #5
            \end{array}$
            \settowidth{\fktdefhilfslaenge}{$#1$:}
            \hspace*{0.6 \fktdefhilfslaenge}  
            \hspace*{\fill}}
\newcommand{\fktdefpur}[5]%
           {$\begin{array}[t]{cccc}%
            #1: & #2 & \nach & #3 \\    
                & #4 & \auf  & #5
            \end{array}$}
\newcommand{\fktdefabgesetztpur}[5]%
           {
            
            $\begin{array}[t]{cccc}%
            #1: & #2 & \nach & #3 \\    
                & #4 & \auf  & #5
            \end{array}$
            \settowidth{\fktdefhilfslaenge}{$#1$:}
            \hspace*{0.6 \fktdefhilfslaenge}
            
           }
\newcommand{\fktdefabgesetzt}[5]%
           {
           
            \hspace*{\fill}
            $\begin{array}[t]{cccc}%
            #1: & #2 & \nach & #3 \\    
                & #4 & \auf  & #5
            \end{array}$
            \settowidth{\fktdefhilfslaenge}{$#1$:}
            \hspace*{0.6 \fktdefhilfslaenge}  
            \hspace*{\fill}
            
            }
\newcommand{\ohnefktdefabgesetzt}[4]%
           {      

            \hspace*{\fill}
            $\begin{array}[t]{ccc}%
            #1 & \nach & #2 \\
            #3 & \auf  & #4
            \end{array}$
            \hspace*{\fill}

            }
\newcommand{\doppelohnefktdefabgesetzt}[6]%
           {

            \hspace*{\fill}
            $\begin{array}[t]{ccccc}%
            #1 & \nach & #2 & \nach & #3\\
            #4 & \auf  & #5 & \auf  & #6
            \end{array}$
            \hspace*{\fill}

            }
\newcommand{\anhang}%
           {\appendix
            \sectioninh{Anhang}
            \renewcommand{\Abschnittnummer}{%
                  \ifx\ZaehlerbisEbene\Ebenea{\Alph{section}}%
                  \else{%
                        \ifx\ZaehlerbisEbene\Ebeneb{\Alph{section}.\arabic{subsection}}%
                        \else{\Alph{section}.\arabic{subsection}.\arabic{subsubsection}}%
                        \fi}%
                  \fi}%
            \renewcommand{\Abschnittnummerpunkt}{\Abschnittnummer.}     
            }            
\newcommand{\anhangengl}%
           {\appendix
            \sectioninh{Appendix}
            \renewcommand{\Abschnittnummer}{%
                  \ifx\ZaehlerbisEbene\Ebenea{\Alph{section}}%
                  \else{%
                        \ifx\ZaehlerbisEbene\Ebeneb{\Alph{section}.\arabic{subsection}}%
                        \else{\Alph{section}.\arabic{subsection}.\arabic{subsubsection}}%
                        \fi}%
                  \fi}%
            \renewcommand{\Abschnittnummerpunkt}{\Abschnittnummer.}     
            }

\newcounter{wdhlstufe}
\newcommand{\sectioninh}[1]%
           {\section*{#1}%
            \addcontentsline{toc}{section}{#1}}
\newcommand{\bezeichnung}[3]
           {\begin{Einrueckungpur}{\hbox to 6em{#1}\hbox to 2.4em{\hfill#2}}
            #3
            \end{Einrueckungpur}}

\newcommand{\doppelteinfach}{e}

\newcommand{\ifdoppelt}[1]{\ifthenelse{\equal{\doppelteinfach}{d}}{#1}{}}
\newcommand{\ifeinfach}[1]{\ifthenelse{\equal{\doppelteinfach}{e}}{#1}{}}

\newlength{\querfhilfsl}              %

\newlength{\hll}

\newcommand{\zurueck}{\hspace*{-14pt}}

\DefThmUmgeb{Theorem}{Theorem}{\ZaehlerbisEbene}
\DefThmUmgeb{Definition}{Definition}{\ZaehlerbisEbene}
\DefThmUmgeb{Satz}{Theorem}{\ZaehlerbisEbene}
\DefThmUmgeb{Proposition}{Theorem}{\ZaehlerbisEbene}
\DefThmUmgeb{Lemma}{Theorem}{\ZaehlerbisEbene}
\DefThmUmgeb{Folgerung}{Theorem}{\ZaehlerbisEbene}
\DefThmUmgeb{Corollary}{Theorem}{\ZaehlerbisEbene}
\DefThmUmgeb{Vorschrift}{Definition}{\ZaehlerbisEbene}
\DefThmUmgeb{Construction}{Definition}{\ZaehlerbisEbene}
\DefSThmUmgeb{FormSatz}{Theorem}{\ZaehlerbisEbene}{\glqq Satz\grqq} 
\DefThmUmgeb{Vermutung}{Theorem}{\ZaehlerbisEbene}
\DefThmUmgeb{Conjecture}{Theorem}{\ZaehlerbisEbene}
\DefThmUmgeb{Konvention}{Definition}{\ZaehlerbisEbene}
\DefThmUmgeb{Feststellung}{Theorem}{\ZaehlerbisEbene}
\DefUnterNumThmUmgeb{DefinitionZusatzNum}{Definition}{\ZaehlerbisEbene}{DefZN}{Definition}
\DefBspUmgeb{Beispiel}{Beispiel}{subsubsection}
\DefBspUmgeb{Example}{Example}{subsubsection}
\DefBspUmgeb{Frage}{Frage}{section}
\DefBspUmgeb{Question}{Question}{section}
\DefBspUmgeb{Aufgabe}{Aufgabe}{section}
\DefBspUmgeb{Ziel}{Ziel}{section}
\DefBemUmgeb{Bemerkung}
\DefBemUmgeb{Remark}
\DefSBemUmgeb{OffeneFrage}{Offene Frage}
\newcommand{\bdf}{\begin{Definition}}
\newcommand{\edf}{\end{Definition}}
\newcommand{\bvorsch}{\begin{Vorschrift}}
\newcommand{\evorsch}{\end{Vorschrift}}
\newcommand{\bconst}{\begin{Construction}}
\newcommand{\econst}{\end{Construction}}
\newcommand{\bthm}{\begin{Theorem}}
\newcommand{\ethm}{\end{Theorem}}
\newcommand{\bsatz}{\begin{Satz}}
\newcommand{\esatz}{\end{Satz}}
\newcommand{\bprop}{\begin{Proposition}}
\newcommand{\eprop}{\end{Proposition}}
\newcommand{\blem}{\begin{Lemma}}
\newcommand{\elem}{\end{Lemma}}
\newcommand{\bfolg}{\begin{Folgerung}}
\newcommand{\efolg}{\end{Folgerung}}
\newcommand{\bcorr}{\begin{Corollary}}
\newcommand{\ecorr}{\end{Corollary}}
\newcommand{\bfest}{\begin{Feststellung}}
\newcommand{\efest}{\end{Feststellung}}
\newcommand{\bbew}{\begin{Beweis}}
\newcommand{\ebew}{\end{Beweis}}
\newcommand{\bpf}{\begin{Proof}}
\newcommand{\epf}{\end{Proof}}
\newcommand{\bwnum}{\begin{WeiterNumerierteListe}}
\newcommand{\ewnum}{\end{WeiterNumerierteListe}}
\newcommand{\bdfzn}{\begin{DefinitionZusatzNum}}
\newcommand{\edfzn}{\end{DefinitionZusatzNum}}
\newcommand{\bbem}{\begin{Bemerkung}}
\newcommand{\ebem}{\end{Bemerkung}}
\newcommand{\brem}{\begin{Remark}}
\newcommand{\erem}{\end{Remark}}
\newcommand{\bnum}{\begin{NumerierteListe}}
\newcommand{\enum}{\end{NumerierteListe}}
\newcommand{\bunum}{\begin{UnnumerierteListe}}
\newcommand{\eunum}{\end{UnnumerierteListe}}
\newcommand{\bbsp}{\begin{Beispiel}}
\newcommand{\ebsp}{\end{Beispiel}}
\newcommand{\bex}{\begin{Example}}
\newcommand{\eex}{\end{Example}}
\newcommand{\bfrag}{\begin{Frage}}
\newcommand{\efrag}{\end{Frage}}
\newcommand{\bquest}{\begin{Question}}
\newcommand{\equest}{\end{Question}}
\newcommand{\baufg}{\begin{Aufgabe}}
\newcommand{\eaufg}{\end{Aufgabe}}
\newcommand{\bof}{\begin{OffeneFrage}}
\newcommand{\eof}{\end{OffeneFrage}}
\newcommand{\bverm}{\begin{Vermutung}}
\newcommand{\everm}{\end{Vermutung}}
\newcommand{\bconj}{\begin{Conjecture}}
\newcommand{\econj}{\end{Conjecture}}
\newcommand{\bkonv}{\begin{Konvention}}
\newcommand{\ekonv}{\end{Konvention}}
\newcommand{\bglklein}{\begin{StandardUnnumGleichungKlein}}
\newcommand{\eglklein}{\end{StandardUnnumGleichungKlein}}
\newcommand{\bgl}{\begin{StandardUnnumGleichung}}
\newcommand{\egl}{\end{StandardUnnumGleichung}}
\newcommand{\bglrtext}{\begin{XrelYZNumGleichung}}
\newcommand{\eglrtext}{\end{XrelYZNumGleichung}}
\newcommand{\zgl}{\EineZeileGleichung}
\newcommand{\zglqed}{\EineZeileGleichungqed}

\newcommand{\znumgl}{\EineNumZeileGleichung}

\newcommand{\berlgl}{\begin{StandardUnnumGleichung}}
\newcommand{\eerlgl}{\end{StandardUnnumGleichung}}
\newcommand{\beinrueck}{\begin{Einrueckungpur}} 
\newcommand{\eeinrueck}{\end{Einrueckungpur}}
\newcommand{\beinflist}{\begin{EinfachListe}} 
\newcommand{\eeinflist}{\end{EinfachListe}}
\newcommand{\beq}{\begin{equation}}
\newcommand{\eeq}{\end{equation}}
\newcommand{\bhin}{\begin{Hinrichtung}}
\newcommand{\ehin}{\end{Hinrichtung}}
\newcommand{\brueck}{\begin{Rueckrichtung}}
\newcommand{\erueck}{\end{Rueckrichtung}}
\newcommand{\bvl}{\begin{AutoLabelLaengenListe}{\ListNullAbstaende}}
\newcommand{\evl}{\end{AutoLabelLaengenListe}}
\newcommand{\df}[1]{{\bf #1}}
\newcommand{\zglnum}[2]{\znumgl{#1\label{#2}}}
\chardef\tempcat=\the\catcode`\@
\catcode`\@=11

\def\@gobble#1{}
\def\@testgrave{\`}
\def\@stressit{\futurelet\chartest\@stresschar }

\def\@stresschar#1{%
  \ifx #1y\def\result{\futurelet\chartest\@yligature}%
  \else \ifx #1Y\def\result{\futurelet\chartest\@Yligature}%
  \else \ifx\chartest\@testgrave \def\result{\accent"26 }%
  \else \def\result{\accent"26 #1}%
  \fi \fi \fi
  \result }

\def\@yligature{%
  \ifx a\chartest \def\result{\accent"26 \char"1F \@gobble}%
  \else \ifx u\chartest \def\result{\accent"26 \char"18 \@gobble}%
  \else \def\result{\accent"26 y}%
  \fi \fi
  \result }

\def\@Yligature{%
  \ifx a\chartest \def\result{\accent"26 \char"17 \@gobble}%
  \else \ifx A\chartest \def\result{\accent"26 \char"17 \@gobble}%
  \else \ifx u\chartest \def\result{\accent"26 \char"10 \@gobble}%
  \else \ifx U\chartest \def\result{\accent"26 \char"10 \@gobble}%
  \else \def\result{\accent"26 Y}%
  \fi \fi \fi \fi
  \result }

\def\!{\ifmmode \mskip-\thinmuskip \fi}

\def\cyracc{\chardef\i="10%
  \def\cydot{{\kern0pt}}%
  \def\cprime{\char"7E }\def\Cprime{\char"5E }%
  \def\cdprime{\char"7F }\def\Cdprime{\char"5F }%
  \def\dbar{dj}\def\Dbar{Dj}%
  \def\dz{\char"1E }\def\Dz{\char"16 }%
  \def\dzh{\char"0A }\def\Dzh{\char"02 }%
  \def\'##1{\if c##1\char"0F %
    \else \if C##1\char"07 %
    \else \accent"26 ##1\fi \fi }%
  \def\`##1{\if e##1\char"0B %
    \else \if E##1\char"03 %
    \else \errmessage{accent \string\` not defined in cyrillic}%
        ##1\fi \fi }%
  \def\=##1{\if e##1\char"0D %
    \else \if E##1\char"05 %
    \else \if \i##1\char"0C %
    \else \if I##1\char"04 %
    \else \errmessage{accent \string\= not defined in cyrillic}%
        ##1\fi \fi \fi \fi }%
  \def\u##1{\if \i##1\accent"24 i%
    \else \accent"24 ##1\fi }%
  \def\"##1{\if \i##1\accent"20 \char"3D %
    \else \if I##1\accent"20 \char"04 %
    \else \accent"20 ##1\fi \fi }%
  \def\!{\ifmmode \def\result{\mskip-\thinmuskip}%
    \else \def\result{\@stressit}\fi \result}}

\def\keep@cyracc{\let\cyr=\relax \let\i=\relax
        \let\ubar=\relax \let\cydot=\relax
        \let\cprime=\relax \let\Cprime=\relax
        \let\cdprime=\relax \let\Cdprime=\relax
        \let\dbar=\relax \let\Dbar=\relax
        \let\dz=\relax \let\Dz=\relax
        \let\dzh=\relax \let\Dzh=\relax
        \let\'=\relax \let\`=\relax \let\==\relax
        \let\u=\relax \let\"=\relax \let\!=\relax }

\DeclareFontEncoding{OT2}{\cyracc}{}

\catcode`\@=\tempcat

\begin{document}
\title{On the Gribov Problem for Generalized Connections}
\author{Christian Fleischhack\thanks{
            e-mail: 
            fleisch@itp.uni-leipzig.de 
            {\it or}    
            chfl@mis.mpg.de
            } \\   %
        \\
        {\normalsize\em Institut f\"ur Theoretische Physik}\\[\adressabstand]
        {\normalsize\em Universit\"at Leipzig}\\[\adressabstand]
        {\normalsize\em Augustusplatz 10/11}\\[\adressabstand]
        {\normalsize\em 04109 Leipzig, Germany}
        \\[-5\adressabstand]
        {\normalsize\em Max-Planck-Institut f\"ur Mathematik in den
                        Naturwissenschaften}\\[\adressabstand]
        {\normalsize\em Inselstra\ss e 22-26}\\[\adressabstand]
        {\normalsize\em 04103 Leipzig, Germany}}
\date{October 10, 2001}
\maketitle
\begin{abstract}
The bundle structure of the space $\Ab$ of Ashtekar's generalized connections
is investigated in the compact case. 
It is proven that every stratum is a locally trivial fibre bundle.
The only stratum being a principal fibre bundle is the generic stratum.
Its structure group equals the space $\Gb$ of all generalized gauge transforms
modulo the constant center-valued gauge transforms.
For abelian gauge theories the generic stratum is globally trivial and
equals the total space $\Ab$.
However, for a certain class of non-abelian gauge theories -- e.g., all
$SU(N)$ theories -- the generic stratum is nontrivial. This means, there
are no global gauge fixings -- the so-called Gribov problem.
Nevertheless, there is a covering of the generic stratum
by trivializations each having total induced
Haar measure~$1$.
\end{abstract}

{\small
\begin{center}
{\bf Keywords}

\vspace{1ex}

Gribov Problem, Ashtekar Connections, Generic Connections, \\
Gauge Fixing, Ashtekar-Lewandowski Measure
\end{center}
}

{\small
\begin{center}
{\bf PASC}

\vspace{1ex}

11.15.Tk, 02.30.Cj, 02.40.--k
\end{center}
}

{\small
\begin{center}
{\bf MSC 2000}

\vspace{1ex}

81T13 (Primary) 53C05, 55R10, 57R22, 58D19 (Secondary)
\end{center}
}

\newpage

\section{Introduction}
The functional integral approach to quantum field theories
consists of two basic steps: first the determination
of a ``physical'' (Euclidian) measure on the configuration space and second
the reconstruction of the quantum theory via an Osterwalder-Schrader
procedure. In this paper we will focus on a certain issue arising
in the first step -- the Gribov problem.

\leerezeile

The configuration space of smooth gauge field theories has a
very complicated structure. The space $\ag$ of all smooth gauge orbits
is, in general, neither affine, nor compact, nor a manifold.
In contrast, the space $\A$ of all smooth gauge fields
(mathematically, connections) is an affine space,
and therefore (from the mathematical point of view) by far more pleasant.
That is why in the 60s Faddeev and Popov \cite{FP} proposed the following
strategy for the gauge-invariant formulation of functional integrals:
First one defines on $\A$ the formal measure
$\DD A = \prod_{x,\mu} \dd A^\mu(x)$.
Due to the gauge invariance of the Yang-Mills action $S_\YM$ the 
functional integral $\int \e^{\I S_\YM(A)} \DD A$ has to be naively
proportional to the integral $\int \DD g \ident \int \prod_x \dd g(x)$
over all gauge transforms. In order to separate this unphysical 
factor, one chooses a gauge fixing, e.g.\ (as Faddeev and Popov did) 
$\del_\nu A^\nu = 0$, and gets
$\prod_{x,\mu} \dd A_\mu(x) = 
 \fpdet(A) \prod_x \bigl(\delta(\del_\nu A^\nu(x)) \prod_\mu \dd A^\mu(x)\bigr)
           \prod_x \dd g(x)$.
The so called Faddeev-Popov determinant $\fpdet(A)$, 
a kind of a degenerate Jacobi determinant, has now to be chosen 
such that $\fpdet(A) \prod_x \delta(\del_\nu (A\circ g)^\nu(x)) \: \dd g(x)$
becomes a constant independent of $A$.
Finally, 
$\prod_x \bigl(\delta(\del_\nu A^\nu(x))  \prod_\mu \dd A^\mu(x)\bigr)$ 
is to be considered as an equivalent of a measure on $\ag$.

About ten years later, Gribov \cite{f15} noticed a severe 
problem: The Coulomb gauge is (for certain gauge theories)
{\em no}\/ gauge fixing since in some gauge orbits several connections
fulfill the gauge condition. Therefore, besides the 
Faddeev-Popov determinant a corresponding, 
highly nontrivial factor had to be inserted into the functional integral. 
Now, one was hoping to avoid this problem using another gauge fixing.
But this hope broke down a few months later:
Singer \cite{f6} investigated this problem more systematically and
found that the non-existence of gauge fixings is a typical property
of gauge theories with non-compact structure group.
Mathematically, this means simply that there is {\em no}\/ global section
in the foliation $\A\nach\ag$ 
of the set of connections over the gauge orbit space.
Moreover, Singer was able to prove that 
(for structure group $\LG=SU(N)$ and spacetime manifold $M = S^4$)
there is no global section even in the principal fibre bundle of all
irreducible connections.

\leerezeile

Most of these problems can be formulated and could arise in a similar way 
also in the Ashtekar framework which differs (in the 
gauge-theoretical sense) from the standard one mainly in the 
usage of distributional connections instead of only smooth ones.
Therefore we want to study in the present paper
whether there is a Gribov problem
for Ashtekar connections as well and, if necessary, study its impact.
Our considerations can be understood as a continuation of the 
investigations started in \cite{paper2+4,paper3}
on the structure of the configuration space of generalized gauge theories.

\subsection{Preliminaries}
Before we outline the paper we recall the basic facts and 
notations from \cite{paper2+4} where the details can be found. 
Other references are \cite{a72,a48,a30}.

Let the ``space-time'' $M$ be an at least two-dimensional manifold, 
$m\in M$ some fixed point
and the structure group $\LG$ be a compact and 
(for technical reasons) connected Lie group.
$\Pf$ denotes the groupoid of all paths in $M$, $\hg$ the group of all
paths starting and ending in $m$.
The set $\Ab$ of generalized connections $\qa$ is defined by 
$\Ab:=\varprojlim_\GR \LG^{\elanz\Edg(\GR)} = \Hom(\Pf,\LG)$. Here $\GR$ runs
over all finite graphs in $M$. $\Edg(\GR)$ is the set of edges in $\GR$,
$\Ver(\GR)$ will be that of all vertices. 
The set $\Gb$ of generalized gauge transforms $\qg$
is $\Gb:=\varprojlim_\GR \LG^{\elanz\Ver(\GR)} = \Maps(M,\LG)$.
It acts continuously on $\Ab$ via 
$h_{\qa\circ\qg}(\gamma) = g_{\gamma(0)}^{-1} h_\qa(\gamma) g_{\gamma(1)}$
where the path $\gamma$ is in $\Pf$ 
and $h_\qa$ is the homomorphism corresponding to $\qa$.
The stabilizer $\bz(\qa)$ of $\qa$ contains exactly those
gauge transforms that fulfill 
$h_\qa(\gamma_x) = g_m^{-1} h_\qa(\gamma_x) g_x$ for all $x\in M$
and whose $m$-component $g_m$ lies in the holonomy centralizer
$Z(\holgr_\qa)$ of $\qa$, i.e.\ the centralizer of the holonomy group of $\qa$. 
Here, for 
all $x$, $\gamma_x$ is some fixed path from $m$ to $x$. We have 
$\bz(\qa)\iso Z(\holgr_\qa)$. Now, the orbit type of $\qa$ is defined to
be the $\Gb$-conjugacy class of $\bz(\qa)$, but equivalently it can be
defined to be the
$\LG$-conjugacy class
$[Z(\holgr_\qa)]$ 
of the holonomy centralizer of $\qa$. This definition will
be used in the following. The types
are partially ordered by the natural inclusion-induced ordering of classes
of subgroups of $\LG$. A stratum $\Abeq t$ is the set of all connections
$\qa\in\Ab$ having type $t$ and the generic stratum $\Abgen$
is the set of all connections
having the maximal orbit type $[Z]$ where $Z\ident Z(\LG)$ is the center
of $\LG$. $\Abgen$ is an open, dense and $\Gb$-invariant subset of $\Ab$.
Furthermore, there is a slice theorem on $\Ab$. 
This means, for every $\qa\in\Ab$
there is a so-called slice $\qS\teilmenge\Ab$ with $\qa\in\qS$ such that:
\bunum
\item
$\qS\circ\Gb$ is an open neighbourhood of $\qa\circ\Gb$ and
\item
there is an equivariant retraction $F:\qS\circ\Gb\nach\qa\circ\Gb$ with
$F^{-1}(\{\qa\})=\qS$.
\eunum
The most important tool for the proof of this theorem has been a so-called
reduction mapping $\varphi_\ga$. Note, due to the compactness of $\LG$,
any centralizer is finitely generated and consequently
there is a finite set $\ga\teilmenge\hg$ 
of paths starting and ending in $m$ such that 
$Z(h_\qa(\ga)) = Z(\holgr_\qa)$. Since $Z(h_\qa(\ga))$ is the orbit type of
$h_\qa(\ga)$ w.r.t.\ the adjoint action of $\LG$ on $\LG^{\elanz\ga}$,
the reduction mapping $\varphi_\ga:\Ab\nach\LG^{\elanz\ga}$ with
$\qa'\auf h_{\qa'}(\ga)$ lifts the slice theorem from $\LG^{\elanz\ga}$ to $\Ab$.
The notion of a reduction mapping will be crucial again in the present paper.

Finally we note that there is a natural measure $\mu_0$ on $\Ab$,
the so-called Ashtekar-Lewandowski measure or induced Haar measure
\cite{paper3,a48}. At the moment, we should only notice that 
$\mu_0(\pi_\GR^{-1}(U)) = \mu_\Haar(U)$
for all graphs $\GR$ 
with corresponding projections $\pi_\GR : \Ab \nach \LG^{\elanz\Edg(\GR)}$,
$\pi_\GR(\qa) = (h_\qa(e_1),\ldots,h_\qa(e_{\elanz\Edg(\GR)}))$, and
all measurable $U \teilmenge \LG^{\elanz\Edg(\GR)}$.
\subsection{Outline of the Paper}
The guideline for our paper will be given by the following five questions.
\bquest
Is $\pi:\Ab\nach\AbGb$ a fibre bundle?
\equest
In general, the answer is negative; at least, if we demand, that all
fibres $\qa\circ\Gb$ have to be isomorphic $\Gb$-spaces. This
required \cite{Bredon} that
all stabilizers are $\Gb$-conjugate. 
But, this implies \cite{paper2+4} that
all Howe subgroups of $\LG$ have to be $\LG$-conjugate.
This, however, is only possible for abelian $\LG$. 

For arbitrary $\LG$ we can expect bundle structures only on subsets of $\Ab$
where all connections have conjugate stabilizers. 
The maximal sets of that kind are exactly the sets
of connections having one and the same gauge orbit type $t$, i.e.\ the 
strata $\Abeq t$.

\bquest
Are the strata $\pi:\Abeq t\nach\AbGbeq t$ locally trivial fibre bundles?
\equest
It is well-known that 
for actions of compact Lie groups on arbitrary (completely regular)
spaces the strata are always fibre bundles. However, the proof of this theorem
does not require the Lie property of the acting group, but the existence
of a slice theorem. In contrast to the former one the latter one is guaranteed
also in our case of interest \cite{paper2+4}.
Hence, we will be able to prove that the
strata are indeed fibre bundles.

The next stronger question is

\bquest
What strata are principal fibre bundles?
\equest
For a fibre bundle to be a principal fibre bundle, 
at least the structure group of that bundle has to equal 
the typical fibre. Since a stratum $\Abeq t$ has (for general reasons)
the typical fibre $\nkla$ and structure group $\nklna$ 
(for some $\qa\in\Abeq t$),
the stabilizer $\bz(\qa)$ has to be a normal subgroup in $\Gb$.
(Here, $N(\bz(\qa))$ denotes the normalizer of $\bz(\qa)$ w.r.t.\ $\Gb$.)
We will show that this is the case in the generic stratum only.

As every other stratum, the generic stratum is locally trivial.
\bquest
Is $\pi:\Abgen\nach\Abgen/\Gb$ even globally trivial?
\equest
For smooth connections -- as mentioned above -- this 
is not the case; this is just the Gribov problem.

In the case of generalized connections the reduction mapping will
play again the most important r\^ole. 
By its means we can reduce the problem of the triviality of the 
generic stratum in $\Ab$ to the corresponding problem
for the generic stratum in $\LG^k$, i.e.\ the set of all 
$\vec g\in\LG^k$ with $Z(\vec g) = Z(\LG)$. 
Namely, by the finiteness lemma for centralizers \cite{paper2+4} we find 
finitely many 
$g_i\in\LG$ with $Z(\vg) \ident Z(\{g_1,\ldots,g_k\}) = Z(\LG)$. 
Now we choose a graph $\GR$ having 
$k$ edges $\alpha_i\in\hg$ and one vertex $m$ and
denote the corresponding reduction mapping $\pi_\GR:\Ab\nach\LG^k$ 
simply by $\varphi$. Since $\varphi$ is surjective even on 
$\Abgen$ \cite{paper2+4}, 
we get the following commutative diagram
\begin{center}
\vspace*{\CDgap}
\begin{minipage}{7.8cm}
\begin{diagram}[labelstyle=\scriptstyle,height=\CDhoehe,l>=3em]
\Abgen                 & \relax\rnachsurj^{\varphi}   & \LG^k             \\
\relax\dnachsurj_{\pi} &                              & \relax\dnachsurj^{\pi_k} \\
\AbGbgen               & \relax\rnachsurj^{[\varphi]} & \LG^k/\Ad
\end{diagram}
\end{minipage}.
\end{center}
One could now conjecture that the generic stratum should be nontrivial
because otherwise one would get a trivialization of
$\pi_k:\LG^k\nach\LG^k/\Ad$, although in general
$\pi_k$ is not a  fibre bundle.
However, this argumentation is incorrect; namely, there is no 
``gauge invariant'' section for $\varphi$, i.e.\ no induced section
for $[\varphi]$.
But, restricting $\LG^k$ in the diagram above to the generic elements
(and modify the remaining three spaces accordingly to the given maps)
this obstruction disappears. Indeed, we will be able to prove
that $\pi:\Abgen\nach\AbGbgen$ is nontrivial, as far as 
$\pi_k:(\LG^k)_\gen\nach(\LG^k)_\gen/\Ad$ is nontrivial.
Moreover, we will explicitely determine a couple of groups (among them
$\LG = SU(N)$) that fulfill the criterion.
Hence we will prove that the Gribov problem appears for generalized
connections again.

In the case of continuous connections the impact of the Gribov problem
to the calculation of functional integrals is enormous.
Since for the integration trivial subbundles are of special interest,
we have the next natural
\bquest
What ``size'' can trivial subbundles of $\Abgen$ have at most?
\equest
Using the reduction mapping above we can show that the triviality of 
$\pi_k$ over $V\teilmenge(\LG^k)_\gen$ 
implies the triviality of $\pi$ over $\varphi^{-1}(V)\teilmenge\Abgen$.
Consequently we get as a by-product that $\pi$ is trivial on the whole 
$\Abgen \ident \Ab$ for abelian $\LG$.
In the general case our task will be the search for domains where
$\pi_k$ is trivial.
The idea for that comes from the fact that a smooth principal fibre bundle
over a contractible manifold is always trivial.
Hence we triangulate $(\LG^k)_\gen/\Ad$, cut out the lower-dimensional
simplices (this is in particular a set of [Lebesgue] measure $0$) 
and get a disjoint union $W$ of contractible
manifolds. The preimage of that union by
$\pi_k\circ\varphi$ yields the desired trivialization
$U\teilmenge\Abgen$ with $\mu_0(U) = \mu_\Haar(\pi_k^{-1}(W)) = 1$.

This way we have found a trivial subbundle arosen from $\Ab$ by cutting out
certain zero subsets. This means that for generalized connections
the Gribov problem is completely irrelevant for the calculation of 
functional integrals supposed the considered measure on $\Ab$ is
absolutely continuous w.r.t.\ the induced Haar measure.

Up to this limitation we get for generalized connections:
\begin{center}
The Gribow problem appears, but it causes no longer a problem.
\end{center}

\leerezeile

In this paper we investigate first the fibre bundle structure of the strata
and prove that the generic stratum is the only stratum having the structure
of a principal fibre bundle. Next, we show that $\LG^k$ for sufficiently
large $k$ is globally trivial up to a zero subset and transfer this
statement to $\Ab$. The proof is quite technical,
although its idea sketched above is rather simple.
Then we prove the triviality of $\Ab$ in the abelian case.
Finally, we state a group-theoretical criterion for the nontriviality
of $\Abgen$ in the nonabelian case and discuss which concrete $\LG$ 
fulfill this criterion.
The example $\LG = SU(2)$ will be described in 
Appendix \ref{app:su2} in detail.

\section{Fibre Bundle Structure of the Strata}
\label{sect:fibrebdl}
\bprop
\label{prop:Abstrata_faserbdl}
Let $t\in\T$ be a gauge orbit type and $\qa\in\Abeq t$ be some connection
of that type. 

Then the stratum $\Abeq t$ is a locally trivial fibre bundle 
over $\AbGbeq t$ with fibre
$\nkla$ and structure group $\nklna$ acting on 
the typical fibre by left translation.
\eprop

\bpf
We know from \cite{paper2+4} that there is a slice theorem on $\Ab$.
Now the assertion follows immediately because
in general every stratum of a space with slice theorem is
such a fibre bundle (cf.\ Appendix \ref{app:bdlstruct}).
\qed
\epf
\brem
The preceding proposition is a further generalization of a result 
for regular connections to the Ashtekar approach.
Already in 1978, Daniel and Viallet \cite{f4} noticed that classically
gauge fixings always exist at least locally.
This result was confirmed by the later statements 
\cite{f5,f1,f11,f12} 
on the triviality of the strata.
\erem

We know already from \cite{paper2+4} that a lot of structures in $\Gb$
can be reduced to structures in the Lie group $\LG$.
For instance, we have seen that $\bz(\qa)$ is always isomorphic
to $Z(\holgr_\qa)$. This, the other way round, implies that the 
homeomorphy class of each orbit is determined not only by $\nkla$, but also  
by $\nklza$. We will see now that there is an analogous result
for the structure group $\nklna$. 
We start investigating $N(\bz(\qa))$ itself.

\bprop
\label{prop:normalis(bz)}
Let $\qa\in\Ab$ and $\qg\in\Gb$. Furthermore, we again fix  
for every $x\in M$ a path $\gamma_x$ from $m$ to $x$, where
$\gamma_m$ is trivial.

Then we have $\qg\in N(\bz(\qa))$ iff
\bnum{2}
\item
$g_m \in N(Z(\holgr_\qa))$ and
\item
$g_x \in h_\qa(\gamma_x)^{-1} \: Z(Z(\holgr_\qa)) \: g_m \: h_\qa(\gamma_x)$
for all $x\in M$.
\enum
\eprop
\bpf 
In general, 
$\qg\in N(\bz(\qa)) \aequ \qg^{-1} \bz(\qa) \qg \teilmenge \bz(\qa)$.

So let $\qg\in\Gb$ and $\qg'\in\bz(\qa)$. Then we have 
\bgl
 &       & \qg^{-1} \qg' \qg \in \bz(\qa) \\
 & \aequ & \begin{minipage}[t]{0.79\textwidth}
           \bnum{2}
           \item
           $g_m^{-1} \: g'_m \: g_m \in Z(\holgr_\qa)$
           \item
           $h_\qa(\gamma_x) = (g_m^{-1} g'_m g_m)^{-1} \: 
                              h_\qa(\gamma_x) \: 
                              (g_x^{-1} g'_x g_x) \fueralle x\in M$
           \enum
           \end{minipage}\\
 & \aequ & \begin{minipage}[t]{0.79\textwidth}
           \bnum{2}
           \item
           $g_m^{-1} \: g'_m \: g_m \in Z(\holgr_\qa)$
           \item
           $               g_m h_\qa(\gamma_x) g_x^{-1} h_\qa(\gamma_x)^{-1} =
            (g'_m)^{-1} \: g_m h_\qa(\gamma_x) g_x^{-1} h_\qa(\gamma_x)^{-1} \: g'_m
                       \fueralle x\in M$.
           \enum
           \end{minipage} \\
 &       & \erl{since $g'_x = h_\qa(\gamma_x)^{-1} g'_m h_\qa(\gamma_x)$}
\egl           

Hence, 
\bgl
 &       & \qg \in N(\bz(\qa)) \\
 & \aequ & \qg^{-1} \qg' \qg \in \bz(\qa) \fueralle \qg'\in\bz(\qa) \\
 & \aequ & \begin{minipage}[t]{0.79\textwidth}
           \bnum{2}
           \item
           $g_m\in N(Z(\holgr_\qa))$
           \item
           $g_m h_\qa(\gamma_x) g_x^{-1} h_\qa(\gamma_x)^{-1} \in Z(Z(\holgr_\qa))
                       \fueralle x\in M$
           \enum
           \end{minipage}\\
 & \aequ & \begin{minipage}[t]{0.79\textwidth}
           \bnum{2}
           \item
           $g_m\in N(Z(\holgr_\qa))$
           \item
           $g_x\in h_\qa(\gamma_x)^{-1}\: Z(Z(\holgr_\qa)) \: g_m h_\qa(\gamma_x)
                       \fueralle x\in M$.
           \enum
           \end{minipage}
\egl
\qed
\epf

\bex
In the generic stratum we have $Z(\holgr_\qa) = Z(\LG)$, hence
$N(Z(\holgr_\qa)) = \LG$ and $Z(Z(\holgr_\qa)) = \LG$.
Consequently, $N(\bz(\qa)) = \Gb$.
\eex
\bex
In the minimal stratum we have $Z(\holgr_\qa) = \LG$.
Hence $N(Z(\holgr_\qa)) = \LG$ and $Z(Z(\holgr_\qa)) = Z(\LG)$.
By the proposition above we have $\qg\in N(\bz(\qa))$ iff
there is some $g_m\in\LG$ such that 
$g_x \in 
  h_\qa(\gamma_x)^{-1}\: Z(\LG) \: g_m h_\qa(\gamma_x) = 
  Z(\LG) \: h_\qa(\gamma_x)^{-1}\: g_m h_\qa(\gamma_x)$ for all $x\in M$.
\eex
\brem
We realize already at this point that (for nonabelian $\LG$) $\bz(\qa)$
is not a normal subset in $\Gb$ provided $\qa$ has minimal type.
A more general statement can be found in Section \ref{abschn:grib:strat_hfb}.
\erem

The preceding proposition shows how the normalizer of $\bz(\qa)$ 
is determined by objects in the Lie group $\LG$.
\bcorr
\label{corr:iso_normalis}
For all $\qa\in\Ab$ the spaces 
$N(\bz(\qa))$ and $N(Z(\holgr_\qa)) \kreuz \dirprod_{x\neq m} Z(Z(\holgr_\qa))$
are homeomorphic.
\ecorr
\bpf
We see immediately that the map 
$\Psi_1:\qg\auf
        \bigl(g_m,(h_\qa(\gamma_x) g_x h_\qa(\gamma_x)^{-1} g_m^{-1})_{x\neq m}\bigr)$
is a homeomorphism between these two spaces.
\qed
\epf
We emphasize that both subgroups of $\Gb$ are in general
{\em not}\/ isomorphic as topological groups. At least there is 
no ``reasonable" homomorphism. We will discuss this problem a bit more in
detail in Appendix \ref{app:homomorph(normalis)}.
Roughly speaking, 
the homomorphy property is destroyed by the above restriction on $g_x$
to be a (usually non-commutative)
product of $g_m$ with elements in $Z(Z(\holgr_\qa))$.

In order to investigate the structure of $\nklna$, we recall the
form of $\bz(\qa)$. By \cite{paper2+4} 
$\bz(\qa)$ and $Z(\holgr_\qa) \kreuz \dirprod_{x\neq m} \{e_\LG\}$ 
are isomorphic topological groups. Hence, heuristically 
\zgl{\nklna \:\:\: \text{ and } 
      \:\:\: \nklnza \kreuz \dirprod_{x\neq m} Z(Z(\holgr_\qa))}
are homeomorphic.
The group isomorphy, however, is not to be expected:
Since the base centralizer $\bz(\qa)$ 
and the holonomy centralizer $Z(\holgr_\qa)$ are isomorphic
groups, it is unlikely that there arise 
isomorphic groups from originally non-isomorphic groups by factorization.
Indeed there are examples (generic connections for $\LG=SU(2)$)
admitting no such ``reasonable" group isomorphism. 
We will discuss this further in Appendix
\ref{app:homomorph(normalis)}.
\bprop
For every $\qa\in\Ab$
\fktdefabgesetzt{\hspace*{-1.2\tabcolsep}[\Psi_1]}
                {\nklna}
                {\nklnza \kreuz \dirprod_{x\neq m} Z(Z(\holgr_\qa))}
                {[\qg]_{\bz(\qa)}}
                {\bigl([g_m]_{Z(\holgr_\qa)}, 
                       (h_\qa(\gamma_x) g_x 
                        h_\qa(\gamma_x)^{-1} g_m^{-1})_{x\neq m}
                 \bigr)}
is a homeomorphism.
\eprop
\bpf
\bunum
\item
$[\Psi_1]$ is well-defined.

Let $[\qg_1]_{\bz(\qa)} = [\qg_2]_{\bz(\qa)}$, i.e.\
$\qg_1 = \qg' \qg_2$ with $\qg'\in\bz(\qa)$.
Then $g'_m\in Z(\holgr_\qa)$, hence 
$[g_{1,m}]_{Z(\holgr_\qa)} = [g'_m g_{2,m}]_{Z(\holgr_\qa)} 
                           = [g_{2,m}]_{Z(\holgr_\qa)}$.
Moreover,
\bgl
       h_\qa(\gamma_x) g_{1,x} h_\qa(\gamma_x)^{-1} g_{1,m}^{-1} 
 & = & h_\qa(\gamma_x) g'_x g_{2,x} h_\qa(\gamma_x)^{-1} g_{2,m}^{-1} (g'_m)^{-1} \\ 
 & = & h_\qa(\gamma_x) g'_x h_\qa(\gamma_x)^{-1} \: 
       h_\qa(\gamma_x) g_{2,x} h_\qa(\gamma_x)^{-1} g_{2,m}^{-1} \:
       (g'_m)^{-1} \\ 
 & = & g'_m \: 
       h_\qa(\gamma_x) g_{2,x} h_\qa(\gamma_x)^{-1} g_{2,m}^{-1} \:
       (g'_m)^{-1} \\ 
 & = & h_\qa(\gamma_x) g_{2,x} h_\qa(\gamma_x)^{-1} g_{2,m}^{-1}
\egl
due to the properties of $\qg'\in\bz(\qa)$ \cite{paper2+4}
and $\qg_i\in N(\bz(\qa))$. 
\item
$[\Psi_1]$ is surjective and continuous.

Follows immediately from the commutative diagram
\begin{center}
\vspace*{\CDgap}
\begin{minipage}{10.5cm}
\begin{diagram}[labelstyle=\scriptstyle,height=\CDhoehe,l>=3em]
N(\bz(\qa))      & \relax\rnach^{\Psi_1}_{\iso} 
                     & N(Z(\holgr_\qa)) \kreuz \dirprod_{x\neq m} Z(Z(\holgr_\qa)) \\
\relax\dnachsurj &   & \relax\dnachsurj \\
\nklna           & \relax\rnach^{[\Psi_1]}      
                     & \nklnza \kreuz \dirprod_{x\neq m} Z(Z(\holgr_\qa))
\end{diagram}
\end{minipage}
\end{center}
and Corollary \ref{corr:iso_normalis}.
\item
$[\Psi_1]$ is injective.

Let $[\Psi_1]([\qg_1]) = [\Psi_1]([\qg_2])$. Then
$g_{1,m} g_{2,m}^{-1}\in Z(\holgr_\qa)$ 
and 
\zgl{h_\qa(\gamma_x) g_{2,x} h_\qa(\gamma_x)^{-1} g_{2,m}^{-1}
      = h_\qa(\gamma_x) g_{1,x} h_\qa(\gamma_x)^{-1} g_{1,m}^{-1}.}
This implies
\bgl
       g_{2,m} h_\qa(\gamma_x) g_{2,x}^{-1}
 & = & g_{1,m} h_\qa(\gamma_x) g_{1,x}^{-1} \\
 & = & g_{1,m} g_{2,m}^{-1} \: 
       g_{2,m} h_\qa(\gamma_x) g_{2,x}^{-1} h_\qa(\gamma_x)^{-1} \:
       h_\qa(\gamma_x) g_{2,x} g_{1,x}^{-1} \\
 & = & g_{2,m} h_\qa(\gamma_x) g_{2,x}^{-1} h_\qa(\gamma_x)^{-1} \:
       g_{1,m} g_{2,m}^{-1} \: 
       h_\qa(\gamma_x) g_{2,x} g_{1,x}^{-1}
\egl
using $\qg_2\in N(\bz(\qa))$ and $g_{1,m} g_{2,m}^{-1}\in Z(\holgr_\qa)$.
Consequently,
\zgl{h_\qa(\gamma_x) = (g_{1,m} g_{2,m}^{-1})^{-1} 
                        h_\qa(\gamma_x) 
                        g_{1,x} g_{2,x}^{-1},}
thus $\qg_1 \qg_2^{-1} \in \bz(\qa)$.
\item
$[\Psi_1]$ is a homeomorphism.

$N(\bz(\qa))$ is a closed subgroup of $\Gb$, 
hence compact. Thus, $\nklna$ is compact as well,
hence $[\Psi_1]$ is a continuous map of a compact space to a Hausdorff space.
This gives \cite{Kelley} the assertion.
\qed
\eunum
\epf

\section{Modified Principal Fibre Bundles}
Now we will find out which strata are even principal fibre bundles.
Here we will use a slightly modified definition for principal fibre bundles.
Actually, one makes the demand on such a bundle that the structure group
acts freely on the fibres, i.e.\ all stabilizers are to be trivial. 
This will not be the case for generalized connections in general
because every holonomy centralizer -- being isomorphic to the corresponding
stabilizer -- contains at least the center of $\LG$.
Therefore we will factor out exactly these ``bugging" 
parts:
\bdf
A $G$-space $X$ is called \df{principal $G$-fibre bundle}
iff all $x\in X$ have the same stabilizer $S$. 

The \df{structure group} of $\pi$ is $\rnkl{S}{G}$.
\edf
Sometimes we say that also $\pi : X \nach X/G$ is a 
principal $G$-fibre bundle.
\bprop
Let (in the notation above) $X$ be a principal fibre bundle with ``typical''
stabilizer $S$. Then 
$S$ is a normal subgroup in $G$, i.e., $\rnkl{S}{G}$ is a 
topological group.

In a natural manner $\rnkl{S}{G}$ acts continuously and freely on $X$. 
Moreover, $X/G \iso X/(\rnkl S G)$.
\eprop
This way the definition of a (not necessarily locally trivial)
principal fibre bundle above is equivalent to the usual one. 
Here neither the total nor the base space gets changed.
Only the acting group is reduced to its part being essential for the action.
By the proposition above we see that the 
notation ``structure {\em group}'' is reasonable. It coincides with the 
standard definition for fibre bundles. Note, however, that 
the $G$ in a ``principal $G$-fibre bundle'' $X$ does not denote 
the structure group as usual, but the total acting group.
\bpf 
Let $x\in X$ and $g\in G$. Obviously, the
stabilizer of $x\circ g$ equals $G_{x\circ g} = g^{-1} G_x g = g^{-1} S g$. 
Since $X$ is a principal fibre bundle, we have $g^{-1} S g = S$, i.e., $S$ 
is a normal subgroup.

Obviously, the action $x\circ [g]_S := x\circ g$ is well-defined
and continuous. Since from $x \circ [g]_S = x$ we get $x \circ g = x$, 
hence $g\in S$, the action is free. The homeomorphy of the two quotient spaces
is clear as well.
\qed
\epf
Now we are left with the modification of the definition of trivializations.

\bdf
Let $G$ be a compact topological group, 
$X$ be a principal $G$-fibre bundle and $\pi:X\nach X/G$ be the 
canonical projection.
Moreover, let $\mu$ be some normalized measure on $X$.
\bnum{3}
\item
An open $G$-invariant set $U\teilmenge X$ is called \df{local
trivialization} of the principal fibre bundle $X$ iff
there is an equivariant homeomorphism
$\loktriv: U \nach \pi(U) \kreuz (\rnkl S G)$ with
$\pi\einschr{U} = \pr_1 \circ \loktriv$.
Here, ``equivariant'' means
$(\pr_2\circ\loktriv) (x\circ g) = ((\pr_2\circ\loktriv)(x)) \cdot [g]_S$ 
for all $x\in U$ and $g\in G$.

We often call $\loktriv$ itself local trivialization.
\item
A principal fibre bundle is called \df{locally trivial} iff
there is a covering $(U_\iota)_{\iota\in \Iota}$ of $X$ by local 
trivializations $U_\iota$.
\item
A principal fibre bundle is called \df{$\mu$-almost globally trivial} iff
there is a covering $(U_\iota)$ as in the preceding item where additionally
$\mu(U_\iota) = 1$ for all $\iota$.%

In the following we usually say simply 
``almost global'' instead of ``$\mu$-almost global''
provided the measure $\mu$ meant is clear.
\item
A principal fibre bundle is called \df{globally trivial} iff
$X$ is a local trivialization.
\enum
\edf
Here again it should be clear that the definitions above are equivalent to the
standard ones. However, we will use these definitions throughout the whole 
paper when we speak about principal fibre bundles and trivializations.

\section{Principal Fibre Bundle Structure of the Strata}
\label{abschn:grib:strat_hfb}
In order to decide which strata are principal fiber bundles 
we have to study the form of the stabilizers, i.e.\ of
the base centralizers. 

\bdf
The set $\bzz:=\{\qg\in\Gb\mid g_m\in Z(\LG)\text{ and }
                   g_x = g_m \fueralle x\in M\}$
is called \df{base center}.
\edf                                 
\blem
\bnum{3}
\item
The base center is contained in every base centralizer.
\item
A base centralizer is a normal subgroup of $\Gb$ iff
it equals the base center. 
\item
The base centralizer of a connection equals
the base center iff the connection is generic. 
\enum
\elem
\bpf
Let $\qa\in\Ab$ with base centralizer $\bz(\qa)$.
\bnum{3}
\item
We have
\bgl
\qg\in\bz(\qa) & \aequ & \begin{minipage}[t]{0.6\textwidth}
                         \bnum{2}
                         \item
                         $g_m \in Z(\holgr_\qa)$
                         \item
                         $h_\qa(\gamma) = g_m^{-1} h_\qa(\gamma) g_x
                         \fueralle \gamma\in\pf{mx}, x\in M$.
                         \enum
                         \end{minipage}
\egl
Since $Z\teilmenge Z(U)$ for all $U\teilmenge \LG$, we have
$\bzz\teilmenge\bz(\qa)$.
\item
\bhin
Let $\bz(\qa)$ be a normal subgroup in $\Gb$.

Let $\qg\in\bz(\qa)$ and $g_0\in\LG$. Furthermore, let $x\in M$, $x\neq m$,
and $\gamma\in\pf{mx}$ be arbitrary. 
Additionally choose a 
$\qg_0\in\Gb$ with $g_{0,m} = g_0$ and $g_{0,x} = h_\qa(\gamma)$.
Then by assumption
$\qg_0^{-1}\:\qg\:\qg_0\in\bz(\qa)$,
in particular
\bgl 
 h_\qa(\gamma) 
   & = & g_{0,m}^{-1} \: g_m^{-1} \: g_{0,m} \:
         h_\qa(\gamma) \:
         g_{0,x}^{-1} \: g_x \: g_{0,x}\\
   & = & g_0^{-1} \: g_m^{-1} \: g_0 \:
         h_\qa(\gamma) \:
         h_\qa(\gamma)^{-1} \: g_x \: h_\qa(\gamma)\\
   & = & g_0^{-1} \: g_m^{-1} \: g_0 \: g_x \: h_\qa(\gamma),
\egl   
hence $g_m \: g_0 = g_0 \: g_x $.
Since $g_0$ is arbitrary, we have $g_m = g_x$ for all $x\in M$ and
consequently $g_m\in Z$. Thus, $\qg\in\bzz$.

We get $\bz(\qa) \teilmenge \bzz$, hence the equality
by the first part of the present proof.
\ehin
\brueck
Let $\bz(\qa) = \bzz$.

Let now $\qg\in\bzz$ and $\qg_0\in\Gb$.
Then we have $(\qg_0^{-1} \: \qg \: \qg_0)_x = 
           g_{0,x}^{-1} \: g_x \: g_{0,x} =
           g_{0,x}^{-1} \: g_m \: g_{0,x} =
           g_{0,x}^{-1} \: g_{0,x} \: g_m = g_m \in Z$,
hence, in particular, 
$(\qg_0^{-1} \: \qg \: \qg_0)_x = (\qg_0^{-1} \: \qg \: \qg_0)_m$ for all
$x\in M$.
Thus, $\qg_0^{-1} \: \qg \: \qg_0\in\bzz$. Hence,
$\bz(\qa) = \bzz$ is a normal subgroup.
\erueck
\item
\bhin
Let $\qa\nichtin\Abgen$, i.e.\ $Z(\holgr_\qa) \echteobermenge Z$.

Let $g\in Z(\holgr_\qa) \setminus Z$, and set
$\qg:=(h_\qa(\gamma_x)^{-1} \: g \: h_\qa(\gamma_x))_{x\in M}$
with $\gamma_x$ chosen as usual.
Per definitionem we have $\qg\in\bz(\qa)$, but $\qg\nichtin\bzz$.
\ehin
\brueck
Let $\qa\in\Abgen$, i.e.\ $Z(\holgr_\qa) = Z$. 
Now we have
\bgl
\qg\in\bz(\qa) & \aequ & \begin{minipage}[t]{0.55\textwidth}
                         \bnum{2}
                         \item
                         $g_m \in Z(\holgr_\qa) = Z$ 
                         \item
                         $h_\qa(\gamma) = g_m^{-1} h_\qa(\gamma) g_x
                         \fueralle \gamma\in\pf{mx}, x\in M $
                         \enum
                         \end{minipage}\\
               & \aequ & \begin{minipage}[t]{0.55\textwidth}
                         \bnum{2}
                         \item
                         $g_m \in Z$
                         \item
                         $h_\qa(\gamma) = h_\qa(\gamma) g_m^{-1} g_x
                         \fueralle \gamma\in\pf{mx}, x\in M $
                         \enum
                         \end{minipage}\\
               & \aequ & \begin{minipage}[t]{0.55\textwidth}
                         \bnum{2}
                         \item
                         $g_m \in Z$
                         \item
                         $g_m = g_x
                         \fueralle x\in M $
                         \enum
                         \end{minipage}\\
               & \aequ & \qg\in\bzz.
\egl
\qed
\erueck
\enum
\epf
Consequently, only the generic stratum is a principal fibre bundle:
\bprop
\label{prop:Abgen=HFB}
Let $t\in\T$ be a gauge orbit type. Then we have:

The stratum $\Abeq t$ is a principal fibre bundle iff
$t = t_{\max}$, i.e.\ $\Abeq t = \Abgen$.
\eprop
\bpf
By the definition of the gauge orbit type,
all holonomy centralizers $Z(\holgr_\qa)$ 
occurring in a fixed stratum $\Abeq t$ 
are conjugate. As proven in \cite{paper2+4} the same is true 
for the stabilizers $\bz(\qa)$.

Now $\Abeq t$ is a principal fibre bundle iff all connections
in $\Abeq t$ have the same stabilizer. On the other hand,
this is true iff this stabilizer is a normal subgroup in 
$\Gb$.\footnote{Let $G$ be a group acting on $X$ where
all stabilizers $G_x$, $x\in X$, are conjugate. 
If $X$ is a principal fibre bundle, then $G_x = G_y$ for all $x,y\in X$
and, in particular, $G_x = G_{x\circ g} = g^{-1} G_x g$ for all $g$.
Thus, $G_x$ is a normal subgroup in $G$. Conversely, let all $G_x$ be
normal subgroups. Then, since all stabilizers are conjugate,
we have $G_y = g^{-1} G_x g \teilmenge G_x$ for all $x,y\in X$, i.e., $X$ is a 
principal fibre bundle.}
The lemma above yields the assertion.
\qed
\epf

\section{Almost Global Triviality of $(\LG^k)_\genueberschrift$}
In Section \ref{abschn:grib:Abgen_fastglobtriv}
we will prove that $\Abgen$ is not only
a locally trivial, but even an almost globally trivial
principal fibre bundle. As mentioned in the introduction 
we will deduce this by means of the reduction mapping
from the corresponding statement on the 
generic stratum of $\LG^k$ w.r.t.\ the adjoint action
that we are going to deal with in the present section.
\label{abschn:fastglobtriv_LGk}
\bdf
Let $k\in\N_+$.

An element $\vg:=(g_1,\ldots,g_k)\in\LG^k$ (and its orbit, resp.) 
is called \df{generic} iff
$\typ(\vg) \ident [Z(\{g_1,\ldots,g_k\})] = [Z]$.

The set of all generic elements of $\LG^k$ is denoted by
$(\LG^k)_\gen$.
\edf
Since $\typ(\vg)\leq [Z]$ for all $\vg\in\LG^k$, 
every generic orbit is of maximal type. 
Moreover, a Lie group is abelian
iff one (and then each) (nontrivial)
power consists of generic elements only.\footnote{We have
$Z(\vec e_\LG) := Z((e_\LG,\ldots,e_\LG)) = \LG$. 
If $\LG^k$ now consists for $k\in\N_+$ 
of generic elements only, then $Z = Z(\vec e_\LG) = \LG$.
Conversely, if $Z=\LG$, then $\LG \obermenge Z(\vg) \obermenge Z = \LG$,
hence $Z(\vg) = Z$ for all $\vg\in\LG^k$.}
\bprop
\label{prop:gener_strat_LGk}
There is a $k_{\min}\in\N_+$, such that the generic stratum
$(\LG^k)_\gen$ is an open and dense submanifold of $\LG^k$ with full
Haar measure
for all $k\geq k_{\min}$ and is empty for all $k < k_{\min}$.
\eprop
\bpf
\bnum{5}
\item
Choice of $\kmin$

By the finiteness lemma for centralizers \cite{paper2+4}
there is a $k\in\N_+$, such that 
there is at least one orbit in $\LG^k$ having type $[Z(\LG)]$.
Now, choose for $\kmin$ simply the minimum of all such $k$.
\bunum
\item
Obviously $(\LG^k)_\gen = \leeremenge$ for $k < k_{\min}$.
\item
For $k\geq \kmin$ there is at least one 
generic element in $\LG^k$.

Namely, let $\vg:=(g_1,\ldots,g_{\kmin})\in\LG^{\kmin}$ be a generic
element. Then 
$\vg_k:=(g_1,\ldots,g_{\kmin},e_\LG,\ldots,e_\LG)\in\LG^{k}$ is generic
as well because of 
\bgl{}
 [Z] & \geq &  \typ(\vg_k) 
    \breitrel= [Z(\{g_1,\ldots,g_{\kmin},e_\LG,\ldots,e_\LG\})] \\
     &   =  &  [Z(\{g_1,\ldots,g_{\kmin}\})] 
    \breitrel= \typ(\vg)
    \breitrel= [Z].
\egl
\eunum
Let now $k\geq k_{\min}$ in the following.
\item
$(\LG^k)_\gen$ is a smooth submanifold of $\LG^k$,
since the adjoint action is smooth. \cite{Bredon} 
\item
$(\LG^k)_\gen$ is open, since any of its points
possesses an open neighbourhood whose 
points have at least, hence by the maximality of $[Z]$ even exactly 
type $[Z]$. \cite{Bredon}
\item
Since $\LG$ is connected, $\LG^k/\Ad$ is connected again.
Again by the maximality of $[Z]$,
we have 
$\quer{(\LG^k)_\gen} \ident \quer{(\LG^k)_{Z}}
 = \bigcup_{[\LK]\leq [Z]} (\LG^k)_{\LK} = \LG^k$. \cite{EMS20} 
Here we set
$(\LG^k)_{\LK} = \{\vg\in\LG^k\mid \typ(\vg) = [\LK]\}$. 
\item
Now, $(\LG^k)_{\LH}$ is a smooth submanifold of $\LG^k$
for every closed subgroup $\LH$ of $\LG$ with $[\LH]<[Z]$. \cite{EMS20}
Moreover, 
$\dim (\LG^k)_{\LH} < \dim (\LG^k)_{Z} \ident \dim (\LG^k)_\gen = \dim \LG^k$.
Since the Haar measure of a lower-dimensional submanifold 
vanishes \cite{Dieu3, Dieu4},
we have $\mu_\Haar((\LG^k)_{\LH}) = 0$.
Since there are only finitely many orbit types on $\LG^k$ 
\cite{EMS20}, we get
$\mu_\Haar(\LG^k \setminus (\LG^k)_\gen) 
  = \mu_\Haar(\bigcup_{[\LH] < [Z]} (\LG^k)_{\LH}) = 0$.
\qed
\enum
\epf

\bprop
\label{prop:fastglobtriv_LGk}
$(\LG^k)_\gen$ is a $\mu_\Haar$-almost globally trivial (smooth)
principal fibre bundle having structure group $\nknonabel$
for all $k\geq \kmin$.
\eprop
\bpf
\bnum{5}
\item
Bundle structure over $(\LG^k)_\gen/\Ad$

The adjoint action of $\LG$ on $\LG^k$ is smooth. Hence 
the generic stratum $(\LG^k)_\gen\ident(\LG^k)_{Z}$ is a 
smooth submanifold and
$\pi_k:(\LG^k)_\gen\nach(\LG^k)_\gen/\Ad$ is a smooth
(locally trivial) fibre bundle
with typical fibre and structure group $\nknonabel$
(\cite{Bredon} and note $N(Z) = \LG$).
Since the stabilizer of any point in $(\LG^k)_\gen$ equals $Z$, 
$\pi_k$ is (in our notation) 
even a principal fibre bundle over the generic stratum.
\item
Choice of a neighbourhood $V_{k,\vv}$ for every fixed $\vv\in(\LG^k)_\gen$

First we cut out from the $n$-dimensional smooth manifold
$(\LG^k)_\gen/\Ad$ a neighbourhood $B^n$ of $[\vv]$ diffeomorphic
to an $n$-dimensional ball and get a remaining smooth manifold $N$.
By general arguments \cite{MiStruss,Whitney,Whitehead}
there are a simplicial complex $K$ consisting of countably many simplices
and a smooth triangulation $f:K\nach N$.
Let $K_n$ denote the set of all $n$-cells\footnote{Note that here
(in contrast to the standard definition)
a cell $\dach\sigma$ is the {\em interior}\/ of a simplex $\sigma$. 
Only for dimension $0$ the cell shall be a simplex.} of $K$.
It can be constructed from $K$ by deleting the 
$(n - 1)$-skeleton $K_{\leq n-1}$, i.e.\ all cells
whose dimension is smaller than that of $K$. 
Now define $V_{k,\vv} := \pi_k^{-1}(f(K_n) \cup \inter B^n)$.

\item
Properties of $V_{k,\vv}$
\bunum
\item
$V_{k,\vv}$ is $\Ad$-invariant.
\item
$\vv\in V_{k,\vv}$.
\item
Since $f$ is a smooth triangulation, $f(K_n)$ is a smooth
manifold \cite{Dieu3} that equals the disjoint union of all $f(\dach\sigma)$
where $\sigma$ is an $n$-simplex of $K$.
Since $f$ is a homeomorphism and $\dach\sigma$ always contractible,
also $f(\dach\sigma)$ is always contractible.
The contractiblity of $\inter B^n$ is trivial.
Moreover, $\inter B^n$ and $f(K_n)$ are disjoint.
\item
Hence, as a $\pi_k$-preimage of a submanifold of $(\LG^k)_\gen/\Ad$
the set $V_{k,\vv}$ is a submanifold of $(\LG^k)_\gen$, thus of $\LG^k$ as well.
In particular, $V_{k,\vv}$ is open in $\LG^k$ by the continuity of $\pi_k$.
Thus $\pi_k(V_{k,\vv})$ is a disjoint union of contractible
manifolds.
\item
We have
$V_{k,\vv} = (\LG^k)_\gen \setminus 
       \bigl(\pi_k^{-1}(\del B^n) \cup
             \bigcup_{\sigma\in K_{\leq n-1}} \pi_k^{-1}(f(\dach\sigma))
       \bigr)$,
i.e.\ $V_{k,\vv}$ emerges from $(\LG^k)_\gen$ by deleting the
$\pi_k$-preimages of the boundary of $B^n$ and 
of all images of lower-dimensional skeletons in $(\LG^k)_\gen/\Ad$,
respectively.
\item
We have $\mu_\Haar(V_{k,\vv}) = 1$.

Since $\mu_\Haar((\LG^k)_\gen) = 1$,
it is sufficient to prove that the just eliminated
objects have Haar measure $0$.
\bnum{2}
\item
Let $\sigma\in K_{\leq n-1}$.
Since $f$ is a homeomorphism, we have 
$\dim f(\dach\sigma) = \dim \dach\sigma < n$. 
As above 
$\pi_k^{-1}(f(\dach\sigma))$ is a submanifold of
$(\LG^k)_\gen$, hence also of $\LG^k$,
whose codimension is $n - \dim \dach\sigma >0$.
Thus, this set is a zero set, too.

Together with $K$, obviously also $K_{\leq n-1}$
is a complex  with countably many 
simplices. Hence,
\zgl{\mu_\Haar\Bigl(\bigcup_{\sigma\in K_{\leq n-1}} \pi_k^{-1}(f(\dach\sigma))\Bigr)
    \leq \sum_{\sigma\in K_{\leq n-1}} \mu_\Haar(\pi_k^{-1}(f(\dach\sigma)))
    = 0.}
\item
$\del B^n$ is a smooth submanifold
of $(\LG^k)_\gen/\Ad$ with codimension $1$. Thus, $\pi_k^{-1}(\del B^n)$
is a smooth submanifold of $(\LG^k)_\gen$ with codimension $1$.

Again, $\mu_\Haar(\pi_k^{-1}(\del B^n))=0$.
\enum
\item
$V_{k,\vv}$ is dense in $\LG^k$.
This follows from $\mu_\Haar(V_{k,\vv})=1$
and the strict positivity of the Haar measure
because $V_{k,\vv}$ is open.
\eunum
\item
Triviality of $\pi_k\einschr{V_{k,\vv}}$

$\pi_k\einschr{V_{k,\vv}}:V_{k,\vv}\nach \pi_k(V_{k,\vv})=\inter B^n \cup f(K_n)$ 
is a principal fibre bundle 
over the disjoint union $\inter B^n \cup f(K_n)$ of contractible manifolds
having structure group $\nknonabel$.
Hence \cite{Husemoller}, this bundle is trivial,
i.e., there is an equivariant
homeomorphism $V_{k,\vv} \iso \pi(V_{k,\vv})\kreuz(\nknonabel)$. 
\item
Almost global triviality of $(\LG^k)_\gen$

Obviously $\ueberdv_k := \{V_{k,\vv} \mid \vv\in(\LG^k)_\gen\}$
is a non-empty covering of $(\LG^k)_\gen$
by $\mu_\Haar$-almost global trivializations.
\qed
\enum
\epf

\section{Relations between the Structure Groups}
\label{abschn:grib:struktgr}
We know already that the structure group of the principal fibre bundle
$(\LG^k)_\gen$ equals $\nknonabel$ and that of the principal fibre bundle
$\Abgen$ equals $\nklagen$. In order to lift the
almost global triviality of $(\LG^k)_\gen$ to that of $\Abgen$
in the next section, we have to investigate the relation between
the corresponding structure groups. We need

\bdf
Let $\punkt:(\nknonabel) \kreuz \Gb \nach \nklagen$ be defined by
$[g]_Z\punkt\qg:= [(g\:g_x)_{x\in M}]_\bzz$.
\edf
\blem
\label{lem:punkt}
\bunum
\item 
$\punkt$ is well-defined and continuous.
\item
The restriction of $\punkt$ to 
$(\nknonabel) \kreuz \Gb_0$ is an isomorphism.
\eunum
\elem
We recall $\Gb_0 := \{\qg\in\Gb\mid g_m = e_\LG\}$.
\bpf
\bunum
\item
Let $g_1\aeqrel g_2$, i.e.\ $g_1 = z g_2$
for a $z\in Z$. Then we have 
\bgl
{}
[g_1]_Z\punkt\qg & = & [(g_1\:g_x)_{x\in M}]_\bzz \\
                 & = & [(z\:g_2\:g_x)_{x\in M}]_\bzz\\
                 & = & [\qz\circ(g_2\:g_x)_{x\in M}]_\bzz
                        \erl{$\qz\ident (z)_{x\in M}\in\bzz$}\\
                 & = & [(g_2\:g_x)_{x\in M}]_\bzz\\
                 & = & [g_2]_Z\punkt\qg.
\egl
\item
The continuity of $\punkt$ follows immediately from the
surjectivity and openness of the canonical projection
$\LG\kreuz\Gb\nach(\nknonabel)\kreuz\Gb$, the continuity
of $\LG\kreuz\Gb\nach\Gb$ with $(g,\qg)\auf (g g_x)_{x\in M}$
as well as that of the canonical projection 
$\Gb\nach\nklagen$ and the commutativity of the corresponding
diagram
\begin{center}
\vspace*{\CDgap}
\begin{minipage}{7.8cm}
\begin{diagram}[labelstyle=\scriptstyle,height=\CDhoehe,l>=3em]
\LG\kreuz\Gb           & \relax\rnach          & \Gb             \\
\relax\dnachsurj_{}    &                       & \relax\dnachsurj \\
(\nknonabel)\kreuz\Gb & \relax\rnach^{\punkt} & \nklagen
\end{diagram}
\end{minipage}.
\end{center}
\item
$\punkt$ is injective.

Let $[g_1]_Z\punkt\qg_1 = [g_2]_Z\punkt\qg_2$ with $\qg_1,\qg_2\in\Gb_0$, i.e.\
$g_{1,m} = g_{2,m} = e_\LG$.
Then per def.
$[(g_1 g_{1,x})_{x\in M}]_\bzz = [(g_2 g_{2,x})_{x\in M}]_\bzz$.
Thus, there is a $z\in Z$ with
$g_1 g_{1,x} = z\: g_2 g_{2,x}$ for all $x\in M$.
For $x=m$ we have $g_1 = z g_2$ (and so $[g_1]_Z = [g_2]_Z$) and
consequently $\qg_1 = \qg_2$.
\item
$\punkt$ is surjective.

Let $[\qg]_\bzz$ be given. Choose a representative
$\qg'\in[\qg]_\bzz$.
Set $g'':= g'_m$ and $g''_x := (g'')^{-1} g'_x$.
Then $[g'']_Z\punkt\qg'' = [\qg]_\bzz$.
\qed
\eunum
\epf

The map $\punkt$ is a modification of the isomorphism
$\redisoeo:\Gb_0\kreuz\nklza\nach\nkla$ \cite{paper2+4}, tailored 
to the special case of generic connections.
There we started with the canonical isomorphism
$\phi'\einschr{Z(\holgr_\qa)} = \phi^{-1} : Z(\holgr_\qa) \nach \bz(\qa), \:
g \auf \bigl(h_\qa(\gamma_x)^{-1}\:g\:h_\qa(\gamma_x)\bigr)_{x\in M}$
and extended it in a natural way to the whole $\LG$.
For generic connections,
$\phi'\einschr Z$ now maps $g$ simply on
$\bigl(h_\qa(\gamma_x)^{-1}\:g\:h_\qa(\gamma_x)\bigr)_{x\in M} = 
 (g)_{x\in M}$.
But, therewith we get a second natural extension of
$\phi'\einschr Z$ to the whole $\LG$. 
This is just the definition of $\punkt$ above.

\section{Almost Global Triviality of $\Ab_\genueberschrift$}
\label{abschn:grib:Abgen_fastglobtriv}
We start with
\bprop
$\Abgen$ has the induced Haar measure $1$.
\eprop
\bpf
By Proposition \ref{prop:gener_strat_LGk},
$(\LG^k)_\gen$ has Haar measure $1$ for some 
$k\in\N_+$. Now, let $\GR$ be some graph
with $k$ edges $\alpha_i$ and the vertex $m$. 
Since the corresponding reduction mapping 
$\varphi \ident \pi_\GR : \Ab \nach \LG^k$ 
decreases the type, \cite{paper2+4} we have
$[Z] = \typ(\varphi(\qa)) \leq \typ(\qa) \leq [Z]$ for all 
$\qa\in\varphi^{-1}((\LG^k)_\gen)$. Hence 
$\Abgen\obermenge \varphi^{-1}((\LG^k)_\gen)$. By definition 
of the induced Haar measure, 
\zglqed%
{1 \geq \mu_0(\Abgen) \geq \mu_0(\varphi^{-1}((\LG^k)_\gen)) 
   = \mu_\Haar((\LG^k)_\gen) = 1.}
\epf
The goal of the present section is the proof of the following
\bthm
\label{thm:Abgen_fastglobtriv}
$\Abgen$ is (w.r.t.\ the action of $\Gb$)
a $\mu_0$-almost globally trivial principal fibre bundle
with structure group $\nklagen$.
\ethm

\subsection{Independent Generators of the Holonomy Centralizer}
Our task is to find for all $\qa\in\Abgen$ a neighbourhood
$U_\qa$ that, on the one hand, has the full measure $1$ and, on the other
hand, is a local trivialization of $\Abgen$.
For that purpose we choose for $\qa$ according to the finiteness lemma for 
centralizers \cite{paper2+4}
finitely many paths $\alpha_i$ in $\hg$ with
$Z(h_\qa(\ga)) = Z(\holgr_\qa) = Z$ and denote the corresponding
reduction mapping shortly by $\varphi$.
The most obvious choice of a neighbourhood
of $\qa$ would be $U:=\varphi^{-1}(V)$, where
$V$ is an almost global trivialization of $(\LG^k)_\gen$
from the covering above with $\varphi(\qa) \in V$.
But, it can happen that despite of $\mu_\Haar(V) = 1$ 
the induced Haar measure of $U$ is {\em smaller than}\/ $1$.
This is, in particular, the case if $\varphi$ is not surjective, 
because, e.g., the paths in $\ga$ are not independent or 
one path in $\ga$ occurs twice what is not a priori forbidden. 
That is why we have to guarantee that we can always find an $\ga$ 
fulfilling a certain independency condition.
For that purpose we will need the notion of hyphs \cite{paper3}:
A hyph $\hyph$ is a set of ``independent'' paths. 
For instance, graphs and webs \cite{d3} are special hyphs.
One of the most striking properties of a hyph is that the 
parallel transports can be assigned to the paths independently. 
Moreover, the corresponding projection 
$\pi_\hyph : \Ab \nach \LG^{\elanz\hyph}$, $\qa\nach h_\qa(\hyph)$,
gives $\mu_0(\pi_\hyph^{-1}(W)) = \mu_\Haar(W)$ for all
$W$.

It would be optimal if we were able to show that for every $\qa$ there
is a hyph $\ga$ with $Z(\holgr_\qa) = Z(h_\qa(\ga))$.
However, though we can find -- starting with an arbitrary $\gb$ 
with $Z(\holgr_\qa) = Z(h_\qa(\gb))$ -- a hyph $\hyph$
such that all paths in $\gb$ can be written as products of paths in $\hyph$,
this set $\hyph$ typically consists not only of 
paths in $\hg$, i.e.\ closed paths only.
To avoid this problem we weaken the notion of a hyph.
\bdf
A finite set $\ga\teilmenge\hg$ is called \df{weak hyph}
iff the corresponding reduction mapping
\fktdefabgesetzt{\varphi_\ga}{\Ab}{\LG^{\elanz\ga}}{\qa'}{h_{\qa'}(\ga)}
is surjective and fulfills 
\zgl{\text{$\mu_0(\pi_\ga^{-1}(W)) = \mu_\Haar(W)$ for all
$W \teilmenge \LG^{\elanz\ga}$.}}
\edf
Obviously, every hyph that consists only of closed paths is a weak hyph.
\bprop
\label{prop:ex(weak_hyph=typ)}
For every $\qa\in\Ab$ there is a weak hyph $\ga\teilmenge\hg$ 
with $Z(h_\qa(\ga)) = Z(\holgr_\qa)$.
\eprop
The proof of this proposition is very technical and is therefore 
shifted to Appendix \ref{app:proof(prop:ex(weak_hyph=typ))}.

\subsection{Choice of the Covering of $\Ab_\genueberschrift$}
Now, we are able to define the desired covering of $\Abgen$:
For each $\qa\in\Abgen$ we choose a
weak hyph $\ga\teilmenge\hg$ with $Z(h_\qa(\ga)) = Z(\holgr_\qa) = Z$
by Proposition \ref{prop:ex(weak_hyph=typ)} and denote the
corresponding reduction mapping as usual by $\varphi_\ga$. 
Now we choose according to Section \ref{abschn:fastglobtriv_LGk} 
an almost global trivialization 
$V_{\elanz \ga, \varphi_\ga(\qa)}$ of $(\LG^{\elanz\ga})_\gen$
that contains $\varphi_\ga(\qa)$, and set
$U_\qa := \varphi_\ga^{-1}(V_{\elanz \ga, \varphi_\ga(\qa)})\teilmenge\Ab$.

By $\qa\in U_\qa$ we have
\bprop
$\ueberdabgen:=\{U_\qa\}_{\qa\in\Abgen}$ is a covering of $\Abgen$.
\eprop

\subsection{Properties of the Covering}
We still have to show that each
$U_\qa$ is an almost global trivialization of $\Abgen$.
For this, let us fix an arbitrary $\qa\in\Abgen$ with
reduction mapping $\varphi:=\varphi_\ga$ and set 
simply $k:=\elanz\ga$, $V:=V_{k,\varphi(\qa)}\teilmenge(\LG^k)_\gen$
and $U:=U_\qa=\varphi^{-1}(V)$.

The easily verifiable properties of $U$ are described by 
\blem
\label{lem:eig(ueberdabgen)}
$U$ is an open, dense, $\Gb$-invariant subset 
of $\Abgen$ with $\mu_0(U) = 1$.
\elem
\bpf
By construction, $V$ is an open, dense, $\Ad$-invariant subset 
of $(\LG^k)_\gen$ with $\mu_\Haar(V) = 1$.
Since $\varphi$ always preserves or reduces the types, \cite{paper2+4}
we have $[Z] = \typ(\varphi(\qa')) \leq \typ(\qa') \leq [Z]$, i.e.\
$\qa'\in\Abgen$ for all $\qa'\in U$.
Since $\varphi$ is continuous, $U = \varphi^{-1}(V)$ is open as a subset
of $\Ab$, hence as a subset of $\Abgen$ as well due to the openness of $\Abgen$.
Since $\ga$ is a weak hyph, we have 
$\mu_0(U) = \mu_0(\varphi^{-1}(V)) = \mu_\Haar(V) = 1$
by Proposition \ref{prop:ex(weak_hyph=typ)}.
Due to the strict positivity of $\mu_0$, both statements yield 
the denseness of $U$ in $\Abgen$.
Finally, the $\Gb$-invariance of $U$ follows from the $\Ad$-invariance
of $V$.
\qed
\epf
The most important property of $U$, however, requires a longer proof.
\bprop
$U$ is a local trivialization of $\Abgen$.
\eprop
\bpf 
We denote the equivariant homeomorphism that belongs to 
the almost global trivialization of $\LG^k$ according
Proposition \ref{prop:fastglobtriv_LGk} by 
$\psi:V \nach \pi_k(V)\kreuz(\nknonabel)$. The projection
onto the second component be $\psi_2:V \nach \nknonabel$.
Furthermore, let $\gamma_x$ be for every $x\in M$ some fixed
path from $m$ to $x$. W.l.o.g., $\gamma_m$ is the trivial path.

Now we define the trivialization mapping of $\Ab$:
\fktdefabgesetzt{\Psi}{U}{\pi(U)\kreuz\nklagen.}
                      {h}{([h],\psi_2(\varphi(h))\punkt(h(\gamma_x))_{x\in M})}
\bnum{5}
\item
$\Psi$ is well-defined.

Because of $h\in U$ we have $\varphi(h)\in V$, i.e.\ $\psi_2(\varphi(h))$ is
well-defined.
\item
$\Psi$ is surjective.

Let $([h],[\qg]_\bzz)\in\pi(U)\kreuz\nklagen$ be given. 
By Lemma \ref{lem:punkt} there is exactly one $[g']_Z\in\nknonabel$ and
some $\qg'\in\Gb_0$ with $[g']_Z\punkt\qg' = [\qg]_\bzz$.
Additionally choose some $h'\in[h]$. 
Hence, $h'\in U$ and $\varphi(h') \ident h'(\ga)\in V$.
Let $\vg := \psi^{-1}([h'(\ga)]_\Ad,[g']_Z)$. Since $h'(\ga)$
and $\vg$ are in one and the same orbit w.r.t.\ the adjoint action, there is 
a $\widetilde g\in\LG$ with $\vg = \widetilde g^{-1} h'(\ga) \widetilde g$.
Now, let 
$\quer h(\gamma) := (g'_x)^{-1} \: 
                    \widetilde g^{-1} h'(\gamma_x\gamma\gamma_y^{-1}) \widetilde g \:
                    g'_y$ 
for all $\gamma\in\pf{xy}$.

Obviously, $\quer h$ is gauge equivalent to $h'$ by means of
the gauge transform $(h'(\gamma_x)^{-1} \widetilde g g'_x)_{x\in M}$. Hence 
$[\quer h] = [h'] = [h]$ and $\quer h \in U$.
Moreover,  
$\varphi(\quer h) \ident \quer h(\ga) = \widetilde g^{-1} h'(\ga) \widetilde g$.
Finally, 
\bgl
        \Psi(\quer h) 
  & = & ([\quer h],\psi_2(\varphi(\quer h))\punkt(\quer h(\gamma_x))_{x\in M}) \\
  & = & ([h],\psi_2(\widetilde g^{-1} h'(\ga) \widetilde g)
             \punkt((g'_m)^{-1} \widetilde g^{-1} h'(1) \widetilde g g'_x)_{x\in M}) \\
  & = & ([h],\psi_2(\vg)
             \punkt(g'_x)_{x\in M}) \\
  & = & ([h],[g']_Z\punkt\qg') \\
  & = & ([h],[\qg]_\bzz).
\egl   
\item
$\Psi$ is injective.

Let $\Psi(h_1) = \Psi(h_2)$. 
Then, in particular,
$  \psi_2(\varphi(h_1))\punkt(h_1(\gamma_x))_{x\in M}
 = \psi_2(\varphi(h_2))\punkt(h_2(\gamma_x))_{x\in M}$.
From the bijectivity of $\punkt$ on $(\nknonabel)\kreuz\Gb_0$
we get $h_1(\gamma_x) = h_2(\gamma_x)$ for all $x\in M$ and
$\psi_2(h_1(\ga)) = \psi_2(h_2(\ga))$.
Since by assumption $h_1$ and $h_2$
are gauge equivalent, there is a $\qg\in\Gb$ with
$h_1 = h_2\circ \qg$.
In particular, we have
$h_1(\ga) = g_m^{-1} h_2(\ga)g_m$, i.e., $h_1(\ga)$
and $h_2(\ga)$ are contained in the same orbit.
Due to 
$\psi_2(h_1(\ga)) = \psi_2(h_2(\ga))$ we have 
$h_2(\ga) = h_1(\ga) = g_m^{-1} h_2(\ga) g_m$, i.e.\
$g_m\in Z(h_2(\ga)) = Z$.
Finally we get
$g_x = h_2(\gamma_x)^{-1} g_m h_1(\gamma_x) = h_2(\gamma_x)^{-1} h_1(\gamma_x) g_m
     = g_m$ 
for all $x\in M$, i.e.\ $\qg\in\bzz$.
Due to $h_1, h_2 \in U$, we get $h_1 = h_2 \circ\qg = h_2$. 
\item
$\Psi$ is continuous.

It is sufficient to prove that the projections from $\Psi$ onto the two
factors are continuous:
\bunum
\item
$\Psi_1:=\pr_1\circ\Psi$ is equal to $\pi:U\nach\pi(U)$, hence continuous.
\item
$\Psi_2:=\pr_2\circ\Psi$ is continuous as a concatenation of 
continuous mappings $\varphi$, $\psi_2$, $\pi_{\gamma_x}$ and 
$\punkt$.\footnote{$\pi_{\gamma_x}:\Ab\nach\LG$, $\qa\auf h_\qa(\gamma_x)$.}
\eunum
\item
$\Psi$ is a homeomorphism.\footnote{Note that the standard theorem
on the continuity of the inverse mapping is not applicable because 
$U$ is typically noncompact.}

Since $f$ is bijective and continuous, we only have to show  
(cf.\ Proposition \ref{prop:stetinversschwach} in Appendix \ref{app:topology}), 
that every element 
of $\pi(U)\kreuz\nklagen$ has a compact neighbourhood whose 
preimage is again compact in $U$. 
Let now such an element $([h'],[\qg]_\bzz)$ be given.
Since $U$ is an open subset of the compact Hausdorff space $\Ab$,
$U$ is locally compact. 
Hence, there is a compact neighbourhood
$U'\teilmenge U$ of $h'$. Now, $W:=\pi(U') \kreuz \nklagen$ is the 
desired compact neighbourhood of $([h'],[\qg]_\bzz)$: Due to the 
compactness of $U'$ and $\Gb$, $W$ is compact and by the openness
of $\pi$ also a neighbourhood; moreover, $\pi^{-1}(W) = U' \circ \Gb$
is (again by the compactness of $\Gb$) compact.
\item
$\Psi$ is equivariant.

We have
\bgl
\Psi(h\circ\qg) 
 & = & ([h\circ\qg],
       \psi_2(\varphi(h\circ\qg))\punkt(((h\circ\qg)(\gamma_x))_{x\in M}))\\
 & = & ([h],
       \psi_2(\varphi(h)\circ g_m)\punkt((g_m^{-1}h(\gamma_x) g_x)_{x\in M}))\\
 & = & ([h],
       (\psi_2(\varphi(h))\cdot[g_m]_Z)
        \punkt((g_m^{-1}h(\gamma_x) g_x)_{x\in M}))\\
 &   & \erl{$\cdot$ denotes the multiplication in $\nknonabel$.}\\
 & = & ([h],
       \psi_2(\varphi(h))\punkt(h(\gamma_x) g_x)_{x\in M}) \\
\egl\par\bgl\phantom{\Psi(h\qg)}
 & = & ([h],
        (\psi_2(\varphi(h))\punkt(h(\gamma_x))_{x\in M})\cdot [\qg]_\bzz) \\
 &   & \erl{$\cdot$ now denotes multiplication in $\nklagen$.}\\
 & = & \Psi(h)\circ [\qg]_\bzz. \\
\egl
\qed
\enum
\epf

\subsection{Proof of Theorem \ref{thm:Abgen_fastglobtriv}}
\bpf[Theorem \ref{thm:Abgen_fastglobtriv}]
By Proposition \ref{prop:Abstrata_faserbdl},
$\Abgen$ is (w.r.t.\ the action of $\Gb$) a fibre bundle with
structure group $\nknonabel$ and by Proposition \ref{prop:Abgen=HFB}
even a principal fibre bundle.
The lemmata and propositions of the present section show that
$\ueberdabgen$ is a covering of
$\Abgen$ by almost global trivializations.
\qed
\epf

\section{Triviality of $\Ab$ for Abelian $\LG$}
For commutative structure groups every connection is generic.
Moreover, $\Ab$ is even globally trivial:
\bprop
Let $\LG$ be a commutative compact Lie group.

Then $\Ab$ is a globally trivial principal fibre bundle
with structure group $\nklagenab$.
\eprop
Here, $\Gbconst$ denoted the set of all constant gauge transforms. 
Obviously, $\Gbconst$ equals the base center.
$\nklagenab$
is isomorphic to $\Gb_0$ as a topological group
via $[\qg]_\Gbconst\auf (g_m^{-1} g_x)_{x\in M}$. 
Therefore one can (after an appropriate modification of the 
action) regard $\Ab$ in the abelian case as a principal $\Gb_0$-fibre bundle 
over $\AbGb$.
\bpf
Let
\fktdefabgesetzt{\Psi}{\Ab}{\AbGb\kreuz\nklagenab,}
                      {h}{([h],[(h(\gamma_x))_{x\in M}])}
where $\gamma_x$ is as usual for all $x\in M$
some path from $m$ to $x$ being trivial for $x=m$.

In a commutative group the adjoint
action is trivial, hence
\zgl{\AbGb = \Hom(\hg,\LG)/\Ad = \Hom(\hg,\LG).}
\bnum{4}
\item
$\Psi$ is surjective.

Let $[h]\in\AbGb$ and $[\qg]\in\nklagenab$ be given. As just remarked
there is an $h'\in\Ab$ with $h'\einschr\hg = [h]$.
Now, let $h''(\gamma) := g_x^{-1} h'(\gamma_x\gamma\gamma_y^{-1}) g_y$
for $\gamma\in\pf{xy}$.
Obviously, $h''\in\Ab$ and $\Psi(h'') = ([h],[\qg])$.
\item
$\Psi$ is injective.

Let $h_1, h_2 \in \Ab$ and $\Psi(h_1) = \Psi(h_2)$.
Then $[(h_1(\gamma_x))_{x\in M}] = [(h_2(\gamma_x))_{x\in M}]$,
hence --~$\gamma_m$ is trivial~-- also $h_1(\gamma_x) = h_2(\gamma_x)$ for all
$x\in M$.
The injectivity now follows, because 
two connections are equal if their 
holonomies are equal and if their parallel transports coincide for each $x$ along
at least one path from $m$ to $x$.
\item
$\Psi$ is obviously continuous.
\item
$\Psi^{-1}$ is continuous because $\Ab$ is compact and $\AbGb\kreuz\nklagenab$
is Hausdorff.
\item
$\Psi$ ist clearly equivariant.
\qed
\enum
\epf

\section{Criterion for the Non-Triviality of $\Ab_\genueberschrift$}
\label{sect:grib:nichttriv}
We already know that $\Abgen$ is almost globally trivial and
in the abelian case even globally trivial.
Now we want to know when the generic stratum is nontrivial. 
First we state a sufficient condition for the non-triviality
of $\Abgen$ requiring only a property of $\LG$ and
find then a class of Lie groups having this property. 
Finally we discuss some problems arising 
when we tried to prove
that $\Abgen$ is nontrivial for {\em all}\/ non-commutative $\LG$.

\subsection{General Criterion}
We start with the sufficient condition for the non-triviality
of $\Abgen$.
\bprop
\label{prop:krit_Abgen_nichttriv}
If there is a natural number $k\geq 1$ such that
$(\LG^k)_\gen$ is a nontrivial principal $\LG$-fibre bundle, then
$\Abgen$ (and thus $\Ab$) is nontrivial as well.
\eprop
\bpf
Let $k\in\N_+$ and $(\LG^k)_\gen$ nontrivial.

Suppose there were a section $s:\AbGbgen\nach\Abgen$ for
$\pi:\Abgen\nach\AbGbgen$.
We choose a graph $\GR$ with $k$ edges and exactly one vertex $m$. 
The set of edges is denoted by $\ga\teilmenge\hg$ and 
defines the reduction mapping
$\varphi:=\varphi_\ga$. Moreover, 
we define $U:=\varphi^{-1}((\LG^k)_\gen)$. 
Our goal is now to construct a section $\svpf$ in the bottom line of the 
diagram
\begin{center}
\vspace*{\CDgap}
\begin{minipage}{7.8cm}
\begin{diagram}[labelstyle=\scriptstyle,height=\CDhoehe,l>=3em]
\Abgen    & \relax\lnach^{\obermenge}                      & U
          & \relax\pile{\lnach^{\svp} \\ \rnach_{\varphi}} & (\LG^k)_\gen     \\
\relax\unach^{s}\dnach_{\pi} &            &
            \relax\unach^{s\einschr{[U]}}\dnach_{\pi\einschr{U}} 
          &            & \relax\dnach^{\pi_k}\unach_{s_k} \\
\AbGbgen  & \relax\lnach^{\obermenge}                      & [U] 
          & \relax\pile{\rnach^{[\varphi]}\\ \lnach_{\svpf}} & (\LG^k)_\gen/\Ad
\end{diagram}
\end{minipage}.
\end{center}
We will see that
$\varphi\circ s\circ \svpf$ induces a continuous section for
$\pi_k:(\LG^k)_\gen\nach(\LG^k)_\gen/\Ad$ that, however, does not exist
by assumption.

We construct first a section 
$\svp:(\LG^k)_\gen\nach\Abgen$.
Here we choose for $\svp(\vg)$ that connection, that is build 
by means of the construction method
\cite{paper3} 
out of the trivial connection if one successively assigns the 
components of $\vg$ to the $k$ edges in $\GR$.
Clearly, then $\varphi(\svp(\vg)) = \vg$ for all $\vg\in\LG^k$.
It is easy to see\footnote{Let $e$ be some edge in $M$. There are at most
two indices $i$ and $j$ such that the initial paths of $\alpha_i$ and 
$\alpha_j$ coincide with a partial path of $e$ or $e^{-1}$.
If there were no such indices, then $\pi_e\circ \svp(\vg) = e_\LG$
for all $\vg$. Otherwise $\pi_e\circ \svp(\vg)$ equals
a product of $g_i$, $g_j$, $g_i^{-1}$ or $g_j^{-1}$.
Thus, in any case $\pi_e\circ \svp : \LG^k \nach \LG$ is continuous 
for all edges $e$. Hence, $\svp$ is continuous as well.}
from this method,
that $\svp$ is continuous and obviously maps $(\LG^k)_\gen$ to $\Abgen$.

Now we define for $[\vg]\in(\LG^k)_\gen/\Ad$ a 
mapping $\svpf:(\LG^k)_\gen/\Ad\nach\AbGbgen$ by
$\svpf([\vg]) := \pi(\svp(\vg))$.
\bunum
\item
$\svpf$ is well-defined.

Let $\vg_2 = \vg_1 \circ g$, $g\in \LG$.
Then $\svp(\vg_2) = \svp(\vg_1) \circ \qg$,
where $\qg\in\Gb$ is that gauge transform having the value 
$g$ everywhere. Hence 
$\svpf([\vg_1]) = \svpf([\vg_2])$.
\item
$\svpf$ is a section.

As just proven we have $\svpf\circ\pi_k = \pi\circ \svp$,
thus
\bgl
{}     [\varphi]\circ \svpf 
 & = & [\varphi]\circ \svpf \circ \pi_k \circ (\pi_k)^{-1} 
       \erl{Surjectivity of $\pi_k$}\\
 & = & [\varphi]\circ \pi \circ \svp \circ (\pi_k)^{-1} \\
 & = & \pi_k \circ \varphi \circ \svp \circ (\pi_k)^{-1} 
       \erl{Commutativity of projections}\\
 & = & \pi_k \circ (\pi_k)^{-1}       
       \erl{Section property}\\
 & = & \ido_{(\LG^k)_\gen/\Ad}.       
       \erl{Surjectivity of $\pi_k$}
\egl
\item
$\svpf$ is continuous.

By the quotient criterion, $\svpf$ is continuous iff
$\svpf \circ \pi_k$ is continuous.
But, the latter one is equal to $\pi\circ \svp$, hence continuous.
\eunum
Finally we prove that $s_k := \varphi\circ s\circ \svpf$
is a section for $\pi_k$: We have
\zgl{\pi_k\circ s_k = \pi_k\circ\varphi\circ s\circ \svpf
                = [\varphi]\circ\pi\circ s\circ \svpf
                = \ido_{(\LG^k)_\gen/\Ad},}
because $s$ and $\svpf$ are sections themselves.      
The continuity of $s_k$ is clear. 
Hence, there is a global continuous section in
$\pi_k:(\LG^k)_\gen\nach(\LG^k)_\gen/\Ad$. This is a contradiction
to the assumption that $\pi_k$ is a nontrivial bundle.

Therefore, there is no continuous section over the whole $\AbGbgen$.
\qed
\epf

\subsection{Concrete Criterion}
The crucial question is now what concrete
$\LG$ and $k$ give nontrivial bundles 
$\pi_k:(\LG^k)_\gen\nach(\LG^k)_\gen/\Ad$.
It is quite easy to see (cf.\ Appendix \ref{app:su2})
that in the case of $\LG=SU(2)$ the bundle
is empty for $k=1$, trivial for $k=2$ and nontrivial for $k\geq 3$. 
So maybe typically up to some
$k$ the bundles are trivial, but nontrivial for bigger $k$.
But, is there a $k$ for {\em every}\/ non-abelian $\LG$ such that 
$\pi_k$ is nontrivial?

Up to now, we did not find a complete answer. However, the following two
propositions give a wide class of groups,
for which the bundle is nontrivial starting at some $k$. 
In particular, the proposition above is non-empty,
i.e., its assumptions can be fulfilled.
\bprop
\label{prop:exist(nichttriv_hfb)}
Let $\LG$ be a non-abelian Lie group with 
$\pi_1^\homo(\nknonabel) \neq 1$ and $\pi_1^\homo(\LG) = 1$.%
\footnote{We denote the fundamental group not as usual by
$\pi_1$, but by $\pi_1^\homo$, in order to avoid confusion with
$\pi_k : \LG^k \nach \LG^k/\Ad$.}

Then there is a $k\in\N$ such that $(\LG^k)_\gen$ 
is a nontrivial principal $\LG$-fibre bundle.
\eprop
\bpf
\bunum
\item
Choose $k'\in\N$ so large that $(\LG^{k'})_\gen$ is non-empty 
(cf.\ Proposition \ref{prop:fastglobtriv_LGk}).

By general arguments one sees \cite{Bredon}
that the codimension of all non-generic
strata, i.e.\ all strata whose type is smaller than $[Z]$,
is at least $1$. 

By Corollary \ref{corr:kodim_strata_mult} in 
Appendix \ref{app:kodim_strata_mult}
the non-generic strata in $\LG^{3k'}$ have at least codimension
$3$. Let $k:=3k'$.
\item
Suppose, the principal fibre bundle
$(\LG^k)_\gen$ were trivial.

Then
$(\LG^k)_\gen \iso (\LG^k)_\gen/\Ad \kreuz \nknonabel$,
hence in particular
\zglnum{\pi_1^\homo((\LG^k)_\gen) \iso 
        \pi_1^\homo((\LG^k)_\gen/\Ad) \dirsum 
        \pi_1^\homo(\nknonabel).}{buendelhomotopien}
Since because of the compactness of $\LG$ the number of non-generic
strata in $\LG^k$ is finite \cite{EMS20} and each one of the strata 
is a submanifold of $\LG^k$ \cite{Bredon} 
having codimension bigger or equal $3$, we have 
$\pi_1^\homo((\LG^k)_\gen) \iso \pi_1^\homo(\LG^k)$.
Consequently, \eqref{buendelhomotopien} reduces to 
\zgl{\pi_1^\homo(\LG)^k \iso \pi_1^\homo(\LG^k) \iso 
     \pi_1^\homo((\LG^k)_\gen/\Ad) \dirsum \pi_1^\homo(\nknonabel).}
This, however, is contradiction to the assumptions
$\pi_1^\homo(\LG) = 1$ and 
$\pi_1^\homo(\nknonabel) \neq 1$.

Hence, $(\LG^k)_\gen$ is nontrivial.
\qed
\eunum
\epf

\bprop
\label{prop:list(nontrivLG)}
The assumptions of the proposition above 
are fulfilled, in particular, for all semisimple 
(simply connected) Lie groups whose decomposition into simple
Lie groups contains at least one of the factors
$A_n$, $B_n$, $C_n$, $D_n$, $E_6$ or $E_7$.
\eprop
\bpf
It is well-known that among the simple (and simply connected) compact
Lie groups exactly the representatives of the series listed above 
have nontrivial center. \cite{Helgason} 

The assumption now follows, because for simply connected
$\LG$ the order of $Z$ equals
$\pi_1^\homo(\nknonabel)$ \cite{HilgNeeb}
and the center of the direct product of groups equals the direct product 
of the corresponding centers.
\qed
\epf
In particular, we see that $\Abgen$ is nontrivial for all $\LG = SU(N)$ 
($=A_{N-1}$, $N\geq 2$). 
However, the corresponding problem, e.g., for $\LG = SO(N)$ or
the prominent case $\LG = E_8 \kreuz E_8$ remains unsolved.

We remark that in general for fixed $\LG$ the bundles gets
``more nontrivial" when $k$ increases. Strictly speaking, we have
\bprop
For every $\LG$ the non-triviality of $(\LG^k)_\gen$ 
implies that of $(\LG^{k+1})_\gen$.
\eprop
\bpf
Let $k$ be chosen such that $(\LG^k)_\gen$ is nontrivial. Suppose there
is a section $s_{k+1}$ for 
$\pi_{k+1}:(\LG^{k+1})_\gen \nach (\LG^{k+1})_\gen/\Ad$.

We get the following commutative diagram
\begin{center}
\vspace*{\CDgap}
\begin{minipage}{8.6cm}
\begin{diagram}[labelstyle=\scriptstyle,height=\CDhoehe,l>=3em]
(\LG^{k+1})_\gen   & \relax\lnach^{\obermenge}   & U  
                   & \relax\rnachsurj^{p_k}      & (\LG^k)_\gen             \\
\relax\dnachsurj_{\pi_{k+1}} &              & \relax\dnachsurj_{\pi_{k+1}\einschr{U}} 
                   &                        & \relax\dnachsurj^{\pi_k} \\
(\LG^{k+1})_\gen/\Ad & \relax\lnach^{\obermenge} & [U]  
                   & \relax\rnachsurj^{[p_k]}        & (\LG^k)_\gen/\Ad
\end{diagram}
\end{minipage}.
\end{center}
Here $p_k:(\LG^{k+1})_\gen\nach(\LG^k)_\gen$ is the 
projection onto the first $k$
coordinates and $[p_k]$ is induced in a natural way.
Additionally, we defined $U:=p_k^{-1}((\LG^k)_\gen)$.

Now we reuse the idea for the proof of 
Proposition \ref{prop:krit_Abgen_nichttriv}. First we set 
$i_k:(\LG^k)_\gen\nach U = (\LG^k)_\gen \kreuz \LG$ with $i_k(\vg) := (\vg,e_\LG)$,
where the rhs vector is viewed as an element of
$\LG^{k+1}$. Obviously, $i_k$ is a continuous section for $p_k$,
that additionally -- as can be checked quickly -- defines 
a continuous section for $[p_k]$ 
via $[i_k]([\vg]) := \pi_{k+1}(i_k(\vg))$.
Finally one sees that 
$s_k := p_k \circ s_{k+1}\einschr{[U]} \circ [i_k]$ 
is a section for $\pi_k$. This, however, is a contradiction to
the non-triviality of $\pi_k$.
\qed
\epf

\subsection{Conjecture}
Let us return again to the proof of 
Proposition \ref{prop:krit_Abgen_nichttriv}.
There we deduced via $\varphi$ from the non-triviality of the generic
stratum in $\LG^k$ that the  
preimage $\varphi^{-1}((\LG^k)_\gen)$, hence $\Abgen$ as well
is a nontrivial bundle. But, besides we know that 
$\varphi$ is surjective even as a mapping from $\Abgen$ 
to the whole $\LG^k$. \cite{paper2+4}
It seems to be obvious that one can now deduce from the non-existence 
of a section for $\pi_k$ over the whole space $\LG^k/\Ad$ (and not only 
over the generic stratum as above) analogously to the case above the 
non-existence of a global section in $\Abgen\nach\AbGbgen$. 
But, the existence of a section for the whole $\pi_k$ is rather 
not to be expected because typically in the case of 
non-commutative structure groups $\LG$ 
the mapping $\pi_k$ does not define a fibre bundle. 
(This can easily be seen because 
in $\LG^k$ there occurs both the orbit [i.e.\ the fibre] $\nknonabel$ and
the orbit $\pt = \rnkl\LG\LG$ being never isomorphic.
However, this is not a criterion for the non-existence of a section, but
simply just an indication.) 
Hence, one can guess that 
$\Abgen$ is surely nontrivial for non-commutative $\LG$,
at least as far as $\pi_k$ possesses no section over the whole
$\LG^k/\Ad$.

Unfortunately, we were not able to prove this up to now.
At one point the proof above uses explicitly the fact that only the generic
stratum in $\LG^k$ is considered -- namely, for the definition of $\svpf$. 
A continuation of that mapping from the generic elements of
$(\LG^k)_\gen/\Ad$ to the whole $\LG^k/\Ad$
is not possible as the next proposition shows:
\bprop
\label{prop:nichtexist(schnitt_in_AbGbgen)}
Let $k\in\N_+$ be some number for that  
there are both generic and non-generic elements in $\LG^k$.

Then there is no continuous mapping $\svp:\LG^k\nach\Abgen$, such that
$\svpf:\LG^k/\Ad\nach\AbGbgen$ with
$\svpf([\vg]) := \pi(\svp(\vg))$
is well-defined and that $\varphi\circ \svp = \ido_{\LG^k}$.
\eprop
We need the following
\blem
Let $k,l\in\N_+$ such that there are generic elements in $\LG^l$.

Then we have:
$\LG$ is abelian iff there is a continuous $f:\LG^l\nach\LG^k$ with
\bunum
\item
$\vg_1 \aeqrel \vg_2$ $\impliz$ 
$(\vg_1,f(\vg_1)) \aeqrel (\vg_2,f(\vg_2))$ and
\item
$Z(\vg) \cap Z(f(\vg)) = Z$ for all $\vg\in\LG^l$.
\eunum
\elem
\bpf
\bunum
\item
Let $\LG$ be abelian. Then, e.g., $f(\vg) = e_\LG$, $\vg\in\LG^l$,
fulfills the conditions of the lemma.
\item
Let $\LG$ be non-abelian. Suppose there were such an $f$.
Let $\vg_1,\vg_2\in\LG^l$ be equivalent, i.e.,
let there exist a $g\in \LG$ with $\vg_2 = \vg_1 \circ g$.
By assumption there is also a $g'\in\LG$ with
$(\vg_2,f(\vg_2)) = (\vg_1,f(\vg_1)) \circ g'
                  = (\vg_1\circ g',f(\vg_1)\circ g')$.
Hence, $\vg_1 \circ g = \vg_1 \circ g'$,
i.e.\ $g' = g'' g$ for some $g''\in Z(\vg_1)$.
Consequently, 
$f(\vg_1 \circ g) = f(\vg_2) = f(\vg_1)\circ g'
                  = f(\vg_1)\circ g'' \circ g$.
In particular, for all generic $\vg_1$ we have $g''\in Z$,
i.e.\ $f(\vg_1 \circ g) = f(\vg_1) \circ g$.
Since the generic elements by assumption
form a dense subset in $\LG^l$ (cf.\ Proposition \ref{prop:gener_strat_LGk}) 
and $f$ is to be continuous, 
$f(\vg_1 \circ g) = f(\vg_1) \circ g$ has to hold even for 
all $\vg_1\in\LG^l$ and all $g\in\LG$.
Let $\vg$ now be a non-generic element in $\LG^l$, i.e.,
let there exist a $g\in Z(\vg)\setminus Z$.
But, now
$f(\vg) = f(\vg \circ g) = f(\vg) \circ g$, hence
$g\in Z(f(\vg)) \cap (Z(\vg)\setminus Z) = (Z(\vg) \cap Z(f(\vg)))\setminus Z
= \leeremenge$. 
Therefore all $\vg\in\LG^l$ are generic
in contradiction to the non-commutativity of $\LG$. 
\qed
\eunum
\epf
\bpf[Proposition \ref{prop:nichtexist(schnitt_in_AbGbgen)}]
Suppose there exists such an $\svp:\LG^k\nach\Abgen$.
\bnum{3}
\item
Let $\vg\in\LG^k$ be arbitrary, but fixed.
Due to $\svp(\vg)\in\Abgen$ there is an
$\ga_{\vg}\teilmenge\hg$ with
$h_{\svp(\vg)}(\ga_{\vg})\in(\LG^{\elanz\ga_{\vg}})_\gen$.
Since the generic stratum is always open and since 
together with $\svp$ and $h_{\ga_{\vg}}$ also
$h_{\ga_{\vg}}\circ\svp:\LG^k\nach\LG^{\elanz\ga_{\vg}}$ is continuous,
$U_{\vg} := (h_{\ga_{\vg}}\circ\svp)^{-1}((\LG^{\elanz\ga_{\vg}})_\gen)$
defines an open neighbourhood of $\vg$.
\item
Varying over all $\vg$ one gets an open covering
$\ueberd:=\{U_{\vg}\mid \vg\in\LG^k\}$ of $\LG^k$.
Since $\LG$ is compact, there are finitely many
$\vg_i\in\LG^k$ such that $\bigcup_i U_{\vg_i}=\LG^k$. 
Let now $\ga'$ be the set (the tuple, respectively) of all these
$\ga_{\vg_i}$.
\item
We define $f:=h_{\ga'}\circ \svp:\LG^k\nach\LG^{\elanz\ga'}$
and --~recall $\varphi\ident h_\ga$~--
\zgl{f':=(h_{\ga}\circ\svp,h_{\ga'}\circ\svp)
  \ident(\ido_{\LG^k},f)
   :\LG^k\nach\LG^{k+\elanz\ga'}.}
We have:
\bunum
\item
$f$ is continuous.
\item
Let $\vg',\vg''\in\LG^k$ with $\vg'\aeqrel\vg''$.
From that, due to the assumed well-definedness of $\svpf$,
$\svp(\vg') \aeqrel \svp(\vg'')$ w.r.t.\ $\Gb$. 
Hence, in particular
$(\vg',f(\vg')) = f'(\vg') \aeqrel f'(\vg'') = (\vg'',f(\vg''))$.
\item
$Z(f(\vg)) = Z$ for all $\vg\in\LG^k$.

Let $\vg\in\LG^k$. Then there is an $i$ with $\vg\in U_{\vg_i}$.
Thus, $h_{\ga_{\vg_i}}(\svp(\vg))\in(\LG^{\elanz\ga_{\vg_i}})_\gen$,
hence $f(\vg) = h_{\ga'}\circ\svp(\vg)\in(\LG^{\elanz\ga'})_\gen$ due to
$\ga_{\vg_i}\teilmenge\ga'$.
\eunum
Due to $Z\teilmenge Z(\vg) \cap Z(f(\vg)) \teilmenge Z(f(\vg)) = Z$, 
$f$ fulfills all assumptions of the preceding lemma,
i.e., $\LG$ is abelian in contradiction to the supposition.
\qed
\enum
\epf

Despite these obstacles we close this section 
with an even stronger
\bconj
$\Abgen$ is nontrivial for every non-commutative $\LG$.
\econj
Perhaps, there is even for every non-commutative $\LG$ a $k$ 
such that $(\LG^k)_\gen$ is nontrivial.

\newcommand{\mueins}{\mu_{G,1}}
\newcommand{\muzwei}{\mu_{G,2}}
\newcommand{\munulleins}{\mu_{0,1}}
\newcommand{\munullzwei}{\mu_{0,2}}
\section{Discussion}
We summarize the results of this paper on the structure of 
the generic stratum of $\Ab$, 
noting that the assertions of the following theorem are completely
independent of the base manifold $M$:
\bthm
\label{thm:strukt_gen_strat}
\bnum{4}
\item
$\Abgen$ has the induced Haar measure $1$. 
\item
$\Abgen$ is a $\mu_0$-almost globally trivial principal $\Gb$-fibre bundle
with structure group $\nklagen$.
\item
$\Abgen$ is globally trivial for abelian $\LG$.
\item
$\Abgen$ is not globally trivial for nonabelian, 
simply connected $\LG$ with nontrivial center.
\enum
Here $\bzz\teilmenge\Gb$ is the set of all constant, center-valued
gauge transforms.
\ethm 
Hence, the Gribov problem -- well-known in the case of
regular (Sobolev) connections for a long time -- 
appears in the Ashtekar approach as well.
However, we could refine the statement substantially:
The non-triviality is now concentrated on a zero subset. 
Just this fact has large physical importance -- 
it justifies the definition of the induced Haar measure on $\AbGb$
by the image measure of the corresponding measure on $\Ab$.
Actually, there are two possibilities for the choice 
of such a measure.\footnote{For an earlier discussion see \cite{paper2+4}.}

Let $X$ be some topological space 
equipped with a measure $\mu$ and let $G$ be some topological
group acting on $X$. 
What should the corresponding measure $\mu_G$ on the orbit space $X/G$ 
look like?
On the one hand, one could simply define
$\mueins(U) := \pi_\ast\mu(U) \ident \mu(\pi^{-1}(U))$ for all measurable
$U\teilmenge X/G$, i.e.\ use the image measure
w.r.t.\ the canonical projection $\pi:X\nach X/G$.
On the other hand, one could perform a kind of Faddeev-Popov transformation:
Let us assume that $X$ can be written as a disjoint union of certain sets
$V$ and that each $V$ equals 
$V/G \kreuz \rnkl{G_V}G$ where $G_V$ characterizes 
the type of orbits on $V$. Then one could define the measure of a 
set $U$ that w.l.o.g.\ is contained in one of these $V/G$
naively by
\zglnum{\muzwei(U) := \mu(\pi^{-1}(U)) \: \nu(G_V)}{gl:faddpop}
where 
$\nu$ somehow measures the ``size'' of the stabilizer $G_V$ in $G$.
(The measure of a general $U$ can then be defined via
$\muzwei(U) := \sum_V \muzwei(U\cap (V/G))$.)
In contrast to the first method, here the orbit space and not the total space
is regarded to be primary: If the measure is uniformly distributed
over all points of the total space, the image measure on the orbits 
would in the first case no longer be uniformly distributed; the weighting 
of the orbits comes along according to their size. But, in the second case
the uniform distribution remains, i.e.\ the group degrees of freedom
do not play any r\^ole in Faddeev-Popov context.
That the equation \eqref{gl:faddpop} above can indeed be seen as a 
Faddeev-Popov transformation will be clear by means of a small 
rearrangement. Namely, we have%
\footnote{We drop -- in contrast to the Faddeev-Popov strategy
in the introduction -- the (fixed) measure $\nu'(G)$ of the totally
acting group.}
\zgl{\mu(\pi^{-1}(U)) = \inv{\nu(G_V)} \: \muzwei(U),}
hence
\zglnum{\pi_\ast\mu = \fpdet \kp \muzwei \:\:\:
        \text{ with \fktdef{\fpdet}{X}{\dach\R}{x}{\inv{\nu(\Stab x)}}}.}%
	{gl:faddpop2}
Thus, $\inv{\nu(G_V)} = \Delta$ is nothing but the 
Faddeev-Popov determinant.

But, what does this mean for our concrete problem $X = \Ab$, $G = \Gb$
and $\mu = \mu_0$?
Since the stabilizer of a connection in the generic stratum is minimal,
$\fpdet$ is maximal on $\Abgen$. Moreover, 
$\fpdet$ has to be constant on the whole $\Abgen$. Since we can assume
that $\fpdet$ is not identically $0$, we get the measure 
$\munullzwei$ on $\AbGbgen$ directly from \eqref{gl:faddpop2}. 
Hence, we are left with the nongeneric strata only.
We know already that the stabilizer of a connection is isomorphic to
the corresponding holonomy centralizer. 
But, typically this subgroup of $\LG$ has a higher dimension
than the center has, i.e.\ its measure w.r.t.\ $\nu$ should be infinite.
Hence, $\fpdet$ would vanish on the nongeneric connections, and a complete 
determination of $\munullzwei$ via \eqref{gl:faddpop2}
is no longer possible. 
The simplest way out would be of course to give the nongeneric strata
simply the $\munullzwei$-measure $0$. This is plausible, if one takes
into account that, on the one hand, in the case of (sufficiently smooth)
actions of Lie groups on manifolds the nonmaximal strata 
have lower dimension than the maximal stratum 
has even after being projected down to the orbit space, i.e.\ they can be
regarded as zero subsets, and, on the other hand, the measure on $\AbGb$
could be defined using the projections
$\pi_\hyph : \AbGb\nach\LG^k/\Ad$.
Nevertheless, it remains for the time being unclear, 
whether one excludes this way perhaps physically interesting
phenomena.

However, the danger is relatively small.
The measure $\munullzwei$ is, of course, not yet the 
``dynamical'' measure, but simply a kind of ``kinematical''
measure. This serves only as a -- in a certain sense natural --
background measure by that means using the dynamics of the system
the ``physical'' measure can be constructed.
One could even argue that this kinematical measure should contain
no dynamical information {\em at all}.
Using this point of view -- and using the arguments above
on the measures of nongeneric strata -- we think it is meaningful 
to give these nongeneric strata in $\AbGb$ the induced Haar measure $0$.

Therewith we realize that both possibilities above for defining a measure on
the orbit space are equivalent in the case of $\AbGb$.
Here $\fpdet$ is constant up to a zero subset, hence 
identical $1$, i.e., we have 
$\munulleins = \pi_\ast\mu_0 
   = \fpdet \kp \munullzwei = \munullzwei$.
Additionally, we see that we can restrict ourselves
to an almost global trivialization of $\Abgen$ when integrating.
Consequently, the Gribov problem appearing for certain $\LG$ 
is irrelevant, e.g., for the calculation of expectation values w.r.t.\ 
$\mu_0$.
Moreover: This statement is true even for all measures that are
absolutely continuous w.r.t.\ $\mu_0$.
Therefore, the next step in this context 
should be an investigation of the structure
of physically interesting measures.

\section{Acknowledgements}
The author has been supported in part by the Reimar-L\"ust-Stipendium
of the Max-Planck-Gesellschaft.
The idea of using homotopy arguments for the proof of the non-triviality
on the Lie group level is due to Lorenz Schwachh\"ofer. Moreover, 
the author thanks Detlev Buchholz, Hartmann R\"omer, Gerd Rudolph and
Matthias Schmidt for inspiring discussions.
\anhangengl

\section{Group Isomorphy of the Two Normalizers}
\label{app:homomorph(normalis)}
In Section \ref{sect:fibrebdl} we dealt with structure groups of the strata. 
As in the case of the stabilizer of a connection
we could describe each of these groups using by far simpler 
structures in the underlying Lie group $\LG$. We saw that 
\zgl{\nklna \text{ and } \nklnza\kreuz\dirprod_{x\neq m} Z(Z(\holgr_\qa))}
are homeomorphic. In this appendix we are going to discuss whether
these groups are isomorphic even as topological groups.

In order to reduce the size of the expressions in what follows, we assume
w.l.o.g.\ that the connection $\qa\in\Ab$ being under consideration
has the property that 
$h_\qa(\gamma_x) = e_\LG$ for all $x\in M$ where $\gamma_x$ is as usual
some fixed path from $m$ to $x$ and $\gamma_m$ is trivial. 
This, indeed, is no restriction because 
every $\qa'\in\Ab$ is gauge equivalent to such an $\qa$.
For this, one would simply set 
$\qg:=(h_{\qa'}(\gamma_x)^{-1})_{x\in M}\in\Gb$
and $\qa:=\qa'\circ\qg$.

As we noticed in the main text already, we will restrict ourselves to 
so-called ``reasonable" isomorphisms. 
\bnum{2}
\item
We look only for isomorphisms between 
\zgl{\nklna\text{ and }\nklnza\kreuz\dirprod_{x\neq m} Z(Z(\holgr_\qa))} that 
are induced by an isomorphism
\zgl{\isonorm: N(\bz(\qa)) 
 \nach N(Z(\holgr_\qa))\kreuz\dirprod_{x\neq m} Z(Z(\holgr_\qa))}
between the two non-factorized spaces. This means, 
such a factor isomorphism has to be a continuation of the
natural isomorphism between the base centralizer and the holonomy centralizer,
hence fulfill
$\isonorm(\bz(\qa)) = Z(\holgr_\qa)\kreuz\dirprod_{x\neq m}\{e_\LG\}$.
\item
We denote the mapping, that one gets by concatenation
of $\isonorm$ and of the corresponding projection to the $x$-component
of the image space, by $\isonorm_x$, $x\in M$. 
For a ``reasonable" isomorphism
$\isonorm$ we demand 
\bnum{2}
\item
$\isonorm_m \ident \pi_m : N(\bz(\qa)) \nach N(Z(\holgr_\qa))$ 
and 
\item
$\isonorm_x(\qg)$ depends only on the values of
$\qg$ in $x$ and $m$. 
\enum
\enum
We think viewing at Proposition 
\ref{prop:normalis(bz)} these restrictions are 
natural.
We neglect only ``wild" isomorphisms, i.e.\ mappings that 
mix the points of $M$.

Now, let $\isonorm$ be a ``reasonable" isomorphism.
We fix a point $x\neq m$ and investigate, how the 
projection $\isonorm_x$ of $\isonorm$ to the point $x$ has to look like.

Since $\isonorm$ is to be a ``reasonable" homomorphism,
$\isonorm_x$ is a map from $(g_m, g_x)$ to
$\isonorm_x(\qg) \ident \isonorm_x(g_m,g_x)$.
By Proposition \ref{prop:normalis(bz)}
every $\qg\in N(\bz(\qa))$ is just determined by the values of 
$g_m\in N(Z(\holgr_\qa))$ and of
$z_x\in Z(Z(\holgr_\qa))$: we have $g_x = z_x g_m$.
Hence, $\isonorm_x$ is well-defined iff\footnote{In what follows
we in general drop the index $m$ in $g_m$
and the index $x$ in $z_x$.}
\zglnum{\isonorm_x(g,z g) \in Z(Z(\holgr_\qa)) 
        \text{ for all $g\in N(Z(\holgr_\qa))$ and $z\in Z(Z(\holgr_\qa))$.}}%
        {glanhnorm1}
The homomorphy property implies
$\isonorm(\qg_1) \isonorm(\qg_2) = \isonorm(\qg_1 \qg_2)$ for all
$\qg_i\in N(\bz(\qa))$, hence
\zgl{\isonorm_x(g_1 g_2,z_1 g_1 z_2 g_2) =
     \isonorm_x(g_1,z_1 g_1) \isonorm_x(g_2,z_2 g_2) 
     \text{ for all $g_i\in N(Z(\holgr_\qa))$ and $z_i\in Z(Z(\holgr_\qa))$.}}
\zglnum{}{glanhnorm2}
Now, we define
\zgl{\isonorma(z) := \isonorm_x(1,z) \text{ for $z\in Z(Z(\holgr_\qa))$}} 
and 
\zgl{\isonormb(g) := \isonorm_x(g,g) \text{ for $g\in N(Z(\holgr_\qa))$}.} 
Obviously $\isonorma:Z(Z(\holgr_\qa))\nach Z(Z(\holgr_\qa))$ and
$\isonormb:N(Z(\holgr_\qa))\nach Z(Z(\holgr_\qa))$ are continuous
(hence smooth) homomorphisms and we
have
\zgl{\isonorm_x(g,z g) = \isonorma(z) \isonormb(g).}
$\isonorma$ is even an automorphism of  
$Z(Z(\holgr_\qa))$ because $\isonorma$ is per constr.\ an
injective Lie morphism.\footnote{In general every injective
Lie morphism $f:\LG\nach\LG$ is already a Lie isomorphism,
if $\LG$ is a compact Lie group. This one sees as follows:
$\im f$ is compact as a continuous image of a compact set,
hence closed. Since the image of a homomorphism
in general is a subgroup of the target space, $\im f$ is 
Lie subgroup. By the homomorphism theorem \cite{HilgNeeb} we have
$\im f \iso \LG/\ker f \iso \LG$.
Hence, $\im f$ as a subgroup of $\LG$ has the same dimension and 
the same number of connected components as $\LG$, and thus is equal
to $\LG$. Therefore $f$ is continuous and bijective, hence
a homeomorphism, hence a Lie isomorphism.} 
The injectivity of $\isonorma$ here is a consequence
of our assumption that $\isonorm_m$ is trivial and $\isonorm_x(\qg)$ does not
depend on $g_{x'}$, $x'\neq x,m$.

Now let $g\in N(Z(\holgr_\qa))$ with
$\isonormb(g)\in Z(Z(Z(\holgr_\qa))) = Z(\holgr_\qa)$.
Due to the bijectivity of $\isonorma$ we have 
exactly for those $g$ that
$\isonorm_x(g,zg) = \isonorma(z) \isonormb(g) 
  = \isonormb(g) \isonorma(z) = \isonorm_x(g,gz)$ for 
all $z\in Z(Z(\holgr_\qa))$. This implies that 
$zg = gz$, i.e.\ $g\in Z(Z(Z(\holgr_\qa))) = Z(\holgr_\qa)$.
Hence we get
\zglnum{\isonormb^{-1}(Z(\holgr_\qa)) \teilmenge Z(\holgr_\qa).}{glanh3}
Our assumption 
$\isonorm(\bz(\qa)) = Z(\holgr_\qa)\kreuz\dirprod_{x\neq m}\{e_\LG\}$ 
implies now\footnote{Note that $\bz(\qa)$ due to the special choice
of the $h_\qa(\gamma_x)$ consists just of the constant $Z(\holgr_\qa)$-valued
gauge transforms.} 
$\isonormb(Z(\holgr_\qa)) = \{e_\LG\}$.
This again yields
$\ker\isonormb \teilmenge \isonormb^{-1}(Z(\holgr_\qa)) 
               \teilmenge Z(\holgr_\qa) \teilmenge \ker\isonormb$,
hence 
\zglnum{\ker\isonormb = Z(\holgr_\qa).}{glanh4}
Therefore we have
\zglnum{\nklnza \iso \im \isonormb \teilmenge Z(Z(\holgr_\qa)).}{glanh5}
This, however, cannot always be fulfilled. Let, e.g.,
be $\LG=SU(2)$ and $\qa$ be generic. 
Then $Z(\holgr_\qa) = Z(SU(2)) = \Z_2$ and
$Z(Z(\holgr_\qa)) = N(Z(\holgr_\qa)) = SU(2)$.
We are looking now for a homomorphism
$\isonormb:SU(2)\nach SU(2)$ with $\ker\isonormb = \Z_2$.
By the homomorphism theorem
$SO(3) \iso SU(2)/\Z_2 \iso \im\isonormb$ is a subgroup
of $SU(2)$. This is a contradiction.\footnote{$SO(3)$ as 
(an isomorphic image of) a subgroup of
$SU(2)$ having the identical dimension and the same number of 
connected components has to be even equal $SU(2)$. 
But, this is impossible, because $SO(3)$ and $SU(2)$ are
non-isomorphic.} Hence, in general, there is {\em no}\/ ``reasonable"
isomorphism of topological groups between
$\nklna$ and $\nklnza\kreuz\dirprod_{x\neq m} Z(Z(\holgr_\qa))$.

Finally, we discuss two special cases.
\bunum
\item
$\isonorma(z) = z$ and $\isonormb(g) = g$.

The resulting mapping $\isonorm_x(g,zg) = zg$ just corresponds
to the restriction $\Psi_0$ of the identical map
on $\Gb$. This, however, gives a group isomorphism only if 
$\isonormb$ is indeed a map from $N(Z(\holgr_\qa))$ to
$Z(Z(\holgr_\qa))$, i.e., these two spaces are equal.

This criterion is fulfilled for instance in the generic stratum.
And indeed, in this case $\Psi_0$ is a group isomorphism
between $N(\bz(\qa))$ and 
$N(Z(\holgr_\qa))\kreuz\dirprod_{x\neq m} Z(Z(\holgr_\qa))$. 
Nevertheless, $\Psi_0$ factorizes by condition \eqref{glanh4}
only then to an isomorphism of the 
quotient groups if $Z(\LG) = Z(\holgr_\qa) = \ker\isonormb$ equals $\{e_\LG\}$.

In the minimal stratum $\Psi_0$ is in general no longer an isomorphism because
at least for non-abelian $\LG$ (i.e., if $\Abgen\neq\Abeq{t_{\min}}$),
$N(Z(\holgr_\qa)) = \LG$ is not equal $Z(Z(\holgr_\qa)) = Z(\LG)$.
\item
$\isonorma(z) = z$ and $\isonormb(g) = e_\LG$.

The resulting mapping $\isonorm_x(g,zg) = z$ corresponds here
to the homeomorphism $\Psi_1$ from Corollary \ref{corr:iso_normalis}.
In order to turn $\isonorm$ into a homomorphism,
by condition \eqref{glanh4},
$N(Z(\holgr_\qa)) = \ker \isonormb = Z(\holgr_\qa)$ 
has to hold.

This condition is fulfilled in the minimal stratum. 
(Then, as can be easily checked,
$\Psi_1$ is indeed a group isomorphism.) 

On the other hand, $\Psi_1$ is no isomorphism for generic $\qa$ in the
non-abelian case.
This is clear, because there we have $N(Z(\holgr_\qa)) = \LG$, but 
$Z(\holgr_\qa) = Z(\LG)$.
\eunum
We see again that it is in many cases impossible
to find a group isomorphism that additionally does not depend
explicitly on the respective stratum containing $\qa$.

\section{Stratification of $SU(2)^k$}
\label{app:su2}
Every $SU(2)$-matrix $A$ can be written uniquely as 
$A = 
  \smatrix{ \phantom-a^{\phantom\ast} & b^{\phantom\ast} \\ -b^\ast & a^\ast}$ 
where $a, b\in\C$ are complex numbers fulfilling
${\betrag a}^2 + {\betrag b}^2 = 1$.
Alternatively such a matrix can be seen as a quaternion 
$A = a + b\J \in \H$. In this case we also describe
$A$ by $a_0 + a_1\I + a_2\J + a_3\KK$ or
$a_0 + \vec a$
with $\sum a_i^2=1$, $a_i\in\R$. We have $SU(2)\iso S^3\teilmenge\R^4$.

\subsection{Adjoint Action on $SU(2)$}
Let $A, C\in SU(2)$ with $A = a_0 + a_1\I + a_2\J + a_3\KK$ and
$C = c_0 + c_1\I + c_2\J + c_3\KK$.
It is easy to check that 
the adjoint action in terms of quaternions is
\zgl{C^+ A C = a_0 + \vec a + 
                  2\bigl(\skalprod{\vec a}{\vec c}\vec c    
                         - \skalprod{\vec c}{\vec c}\vec a  
                         + c_0(\vec a \kreuz \vec c)\bigr)
                = a_0 + \vec a +
                  2\bigl(c_0(\vec a \kreuz \vec c) -
                        \vec c\kreuz(\vec a\kreuz\vec c)\bigr).}
Here, $\skalprod\cdot\cdot$ is the canonical scalar product on $\R^3$.
We determine the stabilizer of $A$.

We have $C^{-1} A C = C^+ A C 
                    = A + 
                      2\bigl(\skalprod{\vec a}{\vec c}\vec c
                             - \skalprod{\vec c}{\vec c}\vec a
                             + c_0(\vec a \kreuz \vec c)\bigr)$,
hence 
\znumgl{C\in Z(A)  \:\:\: \aequ \:\:\:  C^{-1} A C = A \:\:\: \aequ \:\:\:
                       \skalprod{\vec a}{\vec c}\vec c
                     - \skalprod{\vec c}{\vec c}\vec a
                     + c_0(\vec a \kreuz \vec c) = 0. \label{eq:stabbedSU2}}
There are two cases:
\bnum{2}
\item
$\vec a = 0$, i.e.\ $A=\pm\EM$. 

Clearly, the rhs of \eqref{eq:stabbedSU2}
is true for all $C\in SU(2)$, i.e.\ $Z(A) = SU(2)$.
\item
$\vec a \neq 0$, i.e.\ $A\neq\pm\EM$. 

Let $C\in Z(A)$. Multiplying the rhs of
\eqref{eq:stabbedSU2} by $\vec a$ we get
$\skalprod{\vec a}{\vec c}\skalprod{\vec c}{\vec a} -
 \skalprod{\vec c}{\vec c}\skalprod{\vec a}{\vec a} = 0$.
This implies due to $\vec a\neq 0$ that
$\vec c = \mu \vec a$ for some $\mu\in\R$.
Conversely, every such $C$ is indeed a solution.
One easily sees $Z(A)\iso U(1)$.
\enum
In the following we interpret a subset $X$ 
of the three-dimensional ball $B^3$ 
also as a subset 
$X := \{A = a_0 + \vec a\in SU(2)\mid\vec a\in X\}$ of $SU(2)$.

\blem
\label{lem:zentralSU2}
For $A = a_0 + \vec a \in SU(2)$ 
we have $Z(A) = \begin{cases}
                          B^3 & \text{ for $a_0^2 = 1$ } \\
             \R\vec a\cap B^3 & \text{ for $a_0^2 \neq 1$ }
             \end{cases}$.
\elem
Since every $SU(2)$-matrix can be diagonalized, there is a diagonal matrix
in every orbit
\bprop
Every orbit $A\circ SU(2)$ w.r.t.\ the adjoint action of $SU(2)$
on itself contains a point of the form
$a_0 + \sqrt{1-a_0^2}\:\I$.
We have $SU(2)/\Ad \iso [-1,1]$ by $[A]\auf \einhalb\tr A$.

The orbits are the small spheres with constant real part $a_0$.
\eprop
\subsection{Adjoint Action on $SU(2)^2$}
A crucial point for the investigation of gauge orbit types
has been the finiteness lemma for 
centralizers \cite{paper2+4}, i.e.\ every centralizer can be represented
as the centralizer of finitely many elements.
When we have dealt with the generic
stratum we have seen that it is important to generate 
this way the center of the structure group.
How many elements do we need at least for that procedure?
\blem
We have $Z = \{\pm\EM\} \ident \Z_2$.
\elem
\bpf
$Z = \bigcap_{A\in SU(2)} Z(A) 
   = B^3 \cap \bigcap_{\vec a\in B^3} \R\vec a
   = \{\vec 0\}
   \ident \{\pm\EM\}$
by Lemma \ref{lem:zentralSU2}.
\qed
\epf
Obviously no single element of $SU(2)$ generates the whole center,
but already two elements are sufficient. 
We only have to guarantee that their centralizers have 
trivial intersection.
\bprop
There are three orbit types on $SU(2)\kreuz SU(2)$.

Explicitly we have for $A, B\in SU(2)$:
\bnum{3}
\item
Type $[SU(2)]$

$Z(A,B) = B^3$ iff $Z(A) = Z(B) = B^3$.
\item
Type $[U(1)]$

$Z(A,B) = \R\vec c \cap B^3$ iff
\bnum{2}
\item
\label{typ2a}
one centralizer equals $B^3$ and one equals $\R\vec c\cap B^3$ or
\item
\label{typ2c}
two centralizers equal $\R\vec c\cap B^3$.
\enum
\item
Type $[\Z_2]$ (generic elements)

$Z(A,B) = \{\vec 0\}$ iff
$Z(A) = \R\vec a\cap B^3$ and
$Z(B) = \R\vec b\cap B^3$ where 
$\vec a$ and $\vec b$ are non-collinear.
\enum
The dimensions of the respective strata are:
\bunum
\item
Type $[SU(2)]$: $0$;
\item
Type $[U(1)]$: $3$ (case \ref{typ2a}) or $4$ (case \ref{typ2c});
\item
Type $[\Z_2]$: $6$.
\eunum
\eprop
Clearly, $(SU(2)^2)_\gen$ is open and dense in $SU(2)^2$.
The orbits in the generic stratum are isomorphic to 
$SU(2)/\Z_2$, i.e.\ are three-dimensional. 
Hence the quotient space $(SU(2)^2)_\gen/\Ad$ is three-dimensional.

Next, we shall find a continuous section in the
generic stratum of $SU(2)^2$. What could be a ``natural"
element describing an orbit $[(A,B)]$?
Let there be given $(A,B)\in (SU(2)^2)_\gen$.
First we are free to use a matrix $C$ for diagonalizing
the first component $A$. We get
$(C^+ A C, C^+ B C)$. 
It remains the freedom to
act with a second matrix $\Delta_\beta$, however, keeping 
$C^+ A C$ invariant. Hence, $\Delta_\beta$ has to be a diagonal matrix.
On the other hand, $\Delta_\beta$ has to transform
the matrix $C^+ B C$. Otherwise, $C^+ B C$ would be a 
diagonal matrix in contradiction to 
$Z(A) \cap Z(B) = \{\vec 0\}$. Hence, by an appropriate choice of $\Delta_\beta$
we can make the secondary diagonal of
$(C\Delta_\beta)^+ B (C\Delta_\beta)$ real. 
Explicitly we get:
\bprop
\label{prop:stdform_SU2}
In every generic orbit there is a unique element of the form
\zgl{\left(\bmat 
                         \lambda &     0        \\
                             0   & \lambda^\ast
           \emat,
           \bmat
                      x          & \sqrt{1-\betrag x{}^2} \\
         -\sqrt{1-\betrag x{}^2} & x^\ast
         
           \emat
     \right)}
where $\betrag\lambda = 1$, $\ima\lambda > 0$ and $\betrag x < 1$. We call
such an element standard form of the orbit (or its elements).

Conversely, every such element defines a unique generic orbit.

Furthermore the mapping $\pi_F:(SU(2)^2)_\gen\nach(SU(2)^2)_\gen$
assigning to every element its standard form is continuous.

Explicitly, we have for the standard form 
of $(A,B) = (a_0 + \vec a, b_0 + \vec b)$:
\bgl
\lambda & = & a_0 + \norm{\vec a} \I \\
     x  & = & b_0 + \frac{\skalprod{\vec a}{\vec b}}{\norm{\vec a}} \I,
\egl
where $\skalprod{\cdot}{\cdot}$ is again the canonical scalar product 
on $\R^3$.
\eprop
On the level of quaternions the element above can be written
as $(\lambda, x + \sqrt{1-\betrag x{}^2}\:\J)$. 
\bpf
\bunum
\item
Existence 

Let $A = a_0 + \vec a = a_0 + a_1\I + a_2\J + a_3\KK \neq \pm\EM$.
Define\footnote{$\delta$ can be chosen arbitrarily with norm $1$ if 
$a_2 = a_3 = 0$.}
\zgl{\varepsilon := \frac{a_1}{\norm{\vec a}} \:\:\:\text{ and } \:\:\:
     \delta := \frac{a_2 + a_3\I}{\sqrt{a_2^{2} + a_3^{2}}}}
as well as
\zgl{C:=\inv{\sqrt 2}
     \bmat
                \I \sqrt{1+\varepsilon}   &   \delta\sqrt{1-\varepsilon}\\
       -\delta^\ast\sqrt{1-\varepsilon}   &   -\I \sqrt{1+\varepsilon}
     \emat.}
One easily checks
$C^+(A,B) C = \left(\smatrix{ 
                         \lambda &     0        \\
                             0   & \lambda^\ast
                       }, \widetilde B
                 \right)$ 
with an appropriate matrix $\widetilde B$,
where $\betrag \lambda = 1$ and $\ima\lambda>0$.\footnote{Note that
$\ima\lambda=0$ is impossible because otherwise $\lambda=\pm 1$,
i.e.\ $A=\pm\EM$.}

Afterwards we choose a $\beta\in\C$ with 
\zgl{\beta^2 = 
  \frac{\widetilde b_2 + \widetilde b_3\I}
       {\sqrt{\widetilde b_2^{2} + \widetilde b_3^{2}}}.}
Here let $\widetilde b_i$ be as usual the quaternionic components of 
$\widetilde B$. Note that the denominator above is always nonzero
because $\widetilde B$ cannot be a diagonal matrix.

Now, on the one hand, the 
diagonal matrix $\Delta_\beta:=\smatrix{\beta & 0 \\ 0 & \beta^\ast}$
commutes with $C^+ A C$ and, on the other hand, the secondary
diagonal in
$\Delta_\beta^+ \widetilde B \Delta_\beta$ 
is real and positive in the upper right corner, thus, in particular, nonzero.
Hence $(C\Delta_\beta)^+ (A,B) (C\Delta_\beta)$ is of the desired type. 

Furthermore, one checks that $\lambda$ and $x$ depend indeed in the 
given manner on $A$ and $B$.\footnote{More general, 
for the action of $C\Delta_\beta$ on a matrix $M = m_0 + \vec m$ we have:
\zgl{(C\Delta_\beta)^+ M (C\Delta_\beta) = 
       m_0 + 
       \frac{\skalprod{\vec a}{\vec m}}{\norm{\vec a}} \:\:\I +
       \frac{\skalprod{\vec a\kreuz\vec b}{\vec a\kreuz\vec m}}
            {\norm{\vec a\kreuz\vec b} \norm{\vec a}} \:\:\J +
       \frac{\skalprod{\vec a\kreuz\vec b}{\vec m}}
            {\norm{\vec a\kreuz\vec b}} \:\:\KK.}
}            
\item
Uniqueness\footnote{Here, all expressions are quaternionic.}

Suppose, there were two such elements
$(A_i, B_i)=(\lambda_i, x_i + \sqrt{1-\betrag{x_i}{}^2}\:\J)$ 
fulfilling the conditions above.
Then there is a $C\in SU(2)$ with $C^+ (A_1, B_1) C = (A_2, B_2)$.
\bunum
\item
$C^+ A_1 C = A_2$

Since conjugate quaternions have always the same real part, we have
$\re \lambda_1 = \re \lambda_2$.
Hence, $\lambda_1 = \lambda_2$ oder $\lambda_1 = \lambda_2^\ast$.
Since $\ima\lambda_i>0$ by assumption, we get
$\lambda_1 = \lambda_2$, thus $A_1 = A_2$.

Moreover, that is why $C$ is in $Z(A_1)$ and equals
$\mu\in\C$ with $\betrag\mu=1$.
\item
$C^+ B_1 C = B_2$

We have
\bgl
x_2 + \sqrt{1-\betrag{x_2}{}^2}\:\J
  & = & B_2 \\
  & = & C^+ B_1 C \\
  & = & \mu^\ast (x_1 + \sqrt{1-\betrag{x_1}{}^2}\:\J) \mu \\
  & = & x_1 + \sqrt{1-\betrag{x_1}{}^2} \:\mu^\ast\mu^\ast\:\J,
\egl
thus first $x_1 = x_2$.
Therefore the expressions containing the roots are equal,
and due to $\betrag{x_i} < 1$ we have
$(\mu^\ast)^2 = 1$, i.e.\ $\mu = \pm 1$.
\eunum
Thus, $C = \pm\EM\in Z$ and hence $(A_1, B_1) = (A_2, B_2)$.
\item
Since we assumed $\betrag x<1$, every element in the proposition above
defines an orbit in the generic stratum.
\item
The continuity of $\pi_F$ is clear because of
$\norm{\vec a}\neq 0$ in the generic stratum.\footnote{But, note that
the map $A\auf C$ itself is {\em not}\/ continuous. 
This can be seen in the special case that $A$ goes to a 
diagonal matrix, i.e.\ for
$a_2,a_3\gegen 0$ and (here) $a_1<0$. Then namely
$\varepsilon\gegen-1$, i.e., $C$ goes as  
$\frac{a_2 + a_3\I}{\sqrt{a_2^{2} + a_3^{2}}}\:\J$,
hence is divergent.}
\qed
\eunum
\epf
We denote the space of all standard forms by $F$.
Then $F$ is homeomorphic to the product of
the upper open semicircle of $U(1)$ ($\lambda$-part) and 
the upper open hemisphere of $S^2$ ($x$-part),
hence is homeomorphic to $\R^3$.
\bprop
We have $F \iso (SU(2)^2)_\gen/\Ad$.
\eprop
\bpf
Let $f:F\nach (SU(2)^2)_\gen/\Ad$ be the concatenation of the embedding
$\iota$ of $F$ into $(SU(2)^2)_\gen$ and the canonical projection
$\pi_2$:
\begin{center}
\vspace*{\CDgap}
\begin{minipage}{7.8cm}
\begin{diagram}[labelstyle=\scriptstyle,height=\CDhoehe,l>=3em]
F                      & \relax\pile{\reinbett^{\iota}\\ 
                                     \lnach_{\pi_F}}  &   (SU(2)^2)_\gen \\
\relax\dnach^{f} & \ldnach_{\pi_2}  & \\
(SU(2)^2)_\gen/\Ad    &   &
\end{diagram}
\end{minipage}
\end{center}
\bunum
\item
$f$ is bijective by Proposition \ref{prop:stdform_SU2}.
\item
$f$ is continuous as a concatenation of continuous maps.
\item
$f^{-1}$ is continuous, because at least locally (around every point)
there is a continuous section $s_2$ for $\pi_2$
such that locally $f^{-1} = \pi_F \circ s_2$ which implies the
local continuity. However, then $f^{-1}$ is globally continuous.
\qed
\eunum

\epf
Consequently, the generic stratum of $SU(2)^2$ is homeomorphic
to $\R^3\kreuz SO(3)$ because $SU(2)/\Z_2 \iso SO(3)$.

We remark that the total orbit space $SU(2)^2/\Ad$ is homeomorphic to
the three-ball $B^3$ where the singular strata are simply its boundary
$S^2$.
\subsection{Adjoint Action on $SU(2)^3$}
We will show here that the adjoint action on $SU(2)^3$ 
leads to a nontrivial generic stratum. The argument here is again a 
pure homotopy argument as in the proof of the more general 
Proposition \ref{prop:exist(nichttriv_hfb)}. 
But, here we will explicitly describe the strata on $SU(2)^3$
and show that the non-generic strata have codimension $4$. 
(By the way, one easily recognizes that the codimension of the 
non-generic strata on $SU(2)^k$ equals $2(k-1)$.)
\bprop  
On $SU(2)\kreuz SU(2)\kreuz SU(2)$ there are three orbit types.

Explicitly we have for $A, B, C\in SU(2)$:
\bnum{3}
\item
Type $[SU(2)]$

(Dimension $0$) $Z(A,B,C) = B^3$ iff $Z(A) = Z(B) = Z(C) = B^3$.
\item
Type $[U(1)]$

$Z(A,B,C) = \R\vec d \cap B^3$ iff 
\bnum{2}
\item
\label{typ32a}
\label{typ32b}
\label{typ32c}
(Dimension $3$) two centralizers equal $B^3$ 
                and one equals $\R\vec d\cap B^3$ or
\item
\label{typ32d}
\label{typ32e}
\label{typ32f}
(Dimension $4$) one centralizer equals $B^3$ and two equal
                $\R\vec d\cap B^3$ or
\item
\label{typ32g}
(Dimension $5$) three centralizers equal $\R\vec d\cap B^3$.
\enum
\item
Type $[\Z_2]$ (generic elements)

$Z(A,B,C) = \{\vec 0\}$ iff
\bnum{2}
\item
\label{typ33a}
\label{typ33b}
\label{typ33c}
(Dimension $6$) one centralizer equals $B^3$ and the remaining two are different
                and not equal $B^3$, or
\item
\label{typ33d}
\label{typ33e}
\label{typ33f}
(Dimension $7$) two centralizers equal $\R\vec d\cap B^3$ and one
                equals $\R\vec a\cap B^3$, but not all are equal, or
\item
\label{typ33g}
(Dimension $9$) all centralizers are different and not equal $B^3$.
\enum
\enum
\eprop
Now we assume that $(SU(2)^3)_\gen$ were trivial. Then
\zgl{(SU(2)^3)_\gen \iso (SU(2)^3)_\gen/\Ad \kreuz \rnkl{\Z_2}{SU(2)},}
hence
\zgl{\pi_1^\homo((SU(2)^3)_\gen) \iso \pi_1^\homo((SU(2)^3)_\gen/\Ad) \kreuz 
\pi_1^\homo(\rnkl{\Z_2}{SU(2)}).}
As we have just seen the codimension of $SU(2)^3\setminus (SU(2)^3)_\gen$ 
equals $4$, i.e., we have 
$\pi_1^\homo((SU(2)^3)_\gen) = \pi_1^\homo(SU(2)^3) = \pi_1^\homo(SU(2))^3 = 1$. 
On the other hand, $\rnkl{\Z_2}{SU(2)} = SO(3)$ and $\pi_1^\homo(SO(3)) = \Z_2$.
Hence,
\zgl{1 \iso \pi_1^\homo((SU(2)^3)_\gen/\Ad) \dirsum \Z_2,}
which obviously is a contradiction.
Hence, $(SU(2)^3)_\gen$ is nontrivial.

\section{Other Kinds of Connections}
\label{appabschn:grib:irred}
In the literatur there are various kinds of 
regular connections characterized again by the holonomy group $\holgr_A$:
\bnum{3}
\item
$\A_\irred := \{A\in\A\mid \holgr_A = \LG\}$ (often called irreducible),
\item
$\A_\gen := \{A\in\A\mid Z(\holgr_A) = Z(\LG)\}$
(often called generic),
\item
$\A_\almgen := \{A\in\A\mid Z(\holgr_A) \text{ is discrete}\}$
(we call it almost generic).
\enum

Unfortunately, the notations are sometimes diverging. 
The corresponding sets fulfill
$\A_\irred\teilmenge\A_\gen\teilmenge\A_\almgen\teilmenge\A$ for 
semi-simple $\LG$.
If $\LG$ is not semi-simple, the
center of $\LG$ is never discrete \cite{BrtD}, hence no centralizer can be discrete.
Thus, $\A_\almgen$ would be empty. However,
the relation between $\A_\irred$ and $\A_\gen$ survives for arbitrary
$\LG$.

As mentioned, e.g., in \cite{f13} the inclusions are not always proper.
Suppose, e.g., $\LG = SU(2)$, then 
$\A_\irred = \A_\almgen$ for simply connected
base manifolds $M$ and
$\A_\almgen = \A$ for certain $M$ depending on the topology
of the bundle $P = P(M,\LG)$. Moreover, $\A_\gen = \A_\almgen$
iff $\LG$ is a product of $SU(N_i)$'s only.
However, in the case of generalized connections the relation
$\Ab_\irred\echteteilmenge\Abgen$ is proper for every (at least one-dimensional,
but not necessarily semi-simple) Lie group $\LG$. This is a simple consequence
the following
\bprop
Let $M$ be at least two-dimensional and $m$ be as always some 
fixed point in $M$.

For every (abstract) subgroup $H\teilmenge\LG$ there is a generalized 
connection whose holonomy group equals $H$.
\eprop
\bpf
Let $\qa_0$ be the trivial connection, i.e.\ we have
$h_{\qa_0}(\gamma) = e_\LG$ for all $\gamma\in\Pf$.
Furthermore, let $m'$ be some point in $M$ different from $m$.
We choose a set $E := \{\gamma_g\}_{g\in H}\teilmenge\Pf$
of edges connecting $m$ and $m'$ that do not intersect each other
(i.e., the only common points of each two edges are their endpoints
$m$ and $m'$, respectively).
Such an $E$ exists always because
$H$ as a subset of $\LG$ has at most the cardinality of $\R$.%
\footnote{Imagine in some chart (w.l.o.g.\ equal $\R^{\dim M}$)
for instance the set $E$ of all field lines 
between a positive charge in $m$ and a negative charge in $m'$. 
Since there are $\aleph_1$ radial directions starting in $m$, 
there are $\aleph_1$ such field lines as well.}
Obviously the set $V_-$ of all starting points is $\{m\}$,
hence finite. By a proposition in \cite{paper3}
there is an $\qa\in\Ab$ with $h_\qa(\gamma_g) = g$ for all $g\in H$.
In particular,
$h_\qa(\gamma_g\: \gamma_{e_\LG}^{-1}) = g$ for all $g\in H$, hence
$H\teilmenge\holgr_\qa$.
On the other hand, by the definition of $\qa$ (cf.\ again \cite{paper3})
and the group property of $H$, we have
$h_\qa(\gamma) \in H$ for all $\gamma\in\Pf$.
Hence, $\holgr_\qa\teilmenge H$.
\qed
\epf
Thus, there are always proper (even countable) 
subgroups $H\echteteilmenge \LG$ with
$Z(H) = Z(\LG)$ and $H = \holgr_\qa$ for some $\qa\in\Ab$. 
Up to now, we do not know whether $\Ab_\irred$ is 
open or closed or whatever in $\Ab_\gen$.
We also do not know whether $\pi:\Ab\nach\AbGb$ may be trivial
on the irreducible connections. This would be an interesting problem,
in particular, because the original paper \cite{f6} by Singer on the Gribov
ambiguity showed the non-triviality just of the bundle of irreducible 
(Sobolev) connections.

\section{Slice Theorem implies Bundle Structure}
\label{app:bdlstruct}
We are going to prove the following
\bprop
\label{prop:strat->fibrebdl}
Let $G$ be a compact topological group and $X$ be a $G$-space.

If a slice theorem holds on $X$, then for every closed subgroup 
$H \teilmenge G$ the stratum $X_{=H} := \{x\in X \mid \Stab x \aeqrel H\}$
is a locally trivial fibre bundle with typical fibre 
$\rnkl H G$ and structure group $\rnkl{H}{N(H)}$ acting on 
$\rnkl H G$ by left translation.
\eprop
Recall that a slice theorem holds on $X$ iff for every $x\in X$ 
there is a ``slice'' $S\teilmenge X$ with $x\in S$ and open saturation
$S \circ G$ and there is an 
equivariant retraction $f$ of $S \circ G$ onto $x \circ G$ with
$f^{-1}(\{x\}) = S$.

For the proof of the proposition above we need 
\blem
\label{lem:aequiv_homeo=N(H)/H}
Let $H$ be a closed subgroup of the compact group $G$.
Then the group of all equivariant homeomorphisms
between $\rnkl H G$ and $\rnkl H G$ is isomorphic to $\rnkl{H}{N(H)}$.
\elem
\bpf
For every such $f:\rnkl H G\nach \rnkl H G$ there is
an $a_f \in G$ with $f(Hg) = H a_f^{-1} g$ for all $g\in G$. \cite{Bredon}
Such an $a_f$ has to fulfill $a_f^{-1} H a_f \teilmenge H$,
hence has to be in $N(H)$. Conversely, every such $a$
determines a unique $f_a$ of the type above.
By $f(H) = H a_f^{-1}$ the map $f\auf a_f$ is unique up to 
multiplication of $a_f$ by elements of $H$. Hence 
$\psi:\Homeo^G (\rnkl H G)\nach \rnkl{H}{N(H)}$, $f\auf [a_f^{-1}]_H$, 
is bijective and by
$\psi(f_1\circ f_2) = \psi(f_1)\psi(f_2)$ 
even a group isomorphism.
\qed
\epf
The proof of Proposition \ref{prop:strat->fibrebdl}
imitates a proof in \cite{Bredon} that was given there in the case of
$G$ being a Lie group. Indeed, this restriction is superfluous
because one needs only the validity of a slice theorem.
\bpf[Proposition \ref{prop:strat->fibrebdl}]
\bunum
\item
Since all strata are invariant under $G$, the validity of a slice theorem
on $X$ implies immediately that on the stratum $X_{=[H]}$
considered as a $G$-space. Hence, we can assume that
all points in $X$ have type $[H]$.
\item
$X$ is a locally trivial fibre bundle.

We construct around every $x\in X$ with
$\Stab x = H$ a bundle chart. (Obviously there is such an $x$ in every orbit.)
For this, let $S$ be a slice through $x$ and 
$f:S\circ G\nach x\circ G$ be the corresponding equivariant 
retraction, and let $s\in S$ and
$h\in G_s:=\{h\in G\mid s\circ h = s\}$. 
$x = f(s) = f(s\circ h) = f(s) \circ h = x \circ h$ implies $h\in H$,
hence $G_s\teilmenge H$. By assumption all stabilizers on $X$
are conjugate, we have $G_s = H$. Thus, $H$ acts trivially on $S$,
hence, $S \circ G\iso S\kreuz_H G \iso S \kreuz \rnkl H G$ 
as $G$-spaces.

Moreover, $S \iso S/H \iso (S\circ G)/G$: Let
$c:S\nach(S\circ G)/G$ with $c(s) = [s]$. 
\bunum
\item
$c$ is surjective by construction.
\item
$c$ is injective, since $c(s_1) = c(s_2)$ implies $s_1\circ g = s_2$ 
for some $g\in G$, hence (using $f$) 
$x\circ g = x$, thus $g\in H = G_s$ and finally $s_2 = s_1\circ g = s_1$.
\item
$c$ is continuous.
\item
$c$ is closed, because first for every closed 
$U\teilmenge S$ 
(by the closedness of $f^{-1}(\{x\})=S$ in $S\circ G$)
$U$ in $S\circ G$, hence (by the compactness of $G$)
also $U\circ G$ in $S\circ G$ is closed, and second
$\pi:S\circ G\nach (S\circ G)/G$ is closed.
\item
$c$ is open, because it is closed and bijective.
\eunum

Using that identification we get a chart map by the following 
commutative diagram:
\begin{center}
\vspace*{\CDgap}
\begin{minipage}{7.8cm}
\begin{diagram}[labelstyle=\scriptstyle,height=\CDhoehe,l>=3em]
\loktriv_S: & & S \circ G          & \relax\rnach^{\iso}   & S \kreuz \rnkl H G \\
        & & \relax\dnach^{\pi} & \relax\ldnach^{\pr_1} &                    \\
        & & S                  &                       &  
\end{diagram}
\end{minipage}.
\end{center}
\item
$\rnkl{H}{N(H)}$ is the structure group of $X$.

Let $\loktriv_S$ and $\loktriv_T$ be two chart mappings as above. 
We define for all $x\in S\cap T$ using
\begin{center}
\vspace*{\CDgap}
\begin{minipage}{10.4cm}
\begin{diagram}[labelstyle=\scriptstyle,height=\CDhoehe,l>=3em]
\loktriv_T\loktriv_S^{-1}:&&(S\cap T)\kreuz \rnkl H G&&\relax\rnach^{\iso}&&(S\cap T)\kreuz \rnkl H G\\
                          &&             & \relax\rdnach&&\relax\ldnach & \\
                          &&             &           &S \cap T&         &
\end{diagram}
\end{minipage}
\end{center}
an equivariant homeomorphism
\zgl{\vartheta_x:\rnkl H G\nach \rnkl H G \text{ by }
\loktriv_T\loktriv_S^{-1}(x, Hg) = (x,\vartheta_x(Hg)).}
According to the proof of Lemma \ref{lem:aequiv_homeo=N(H)/H},
$\vartheta_x$ corresponds via $\vartheta_x(H) = H a_{\vartheta_x}^{-1}$
to a unique $\uebergabb^S_T(x):=[a_{\vartheta_x}]^{-1}\in \rnkl{H}{N(H)}$.
Hence $\loktriv_S^{-1}(x, Hg) = \loktriv_T^{-1}(x,\uebergabb^S_T(x)\circ Hg)$.

We are now left with the proof of the continuity of 
$\uebergabb^S_T:S\cap T\nach \rnkl{H}{N(H)}$.
But, this we get easily from the continuity of 
$\loktriv_T\loktriv_S^{-1}$ by means of 
{

$\begin{array}[t]{ccccccc}%
\hspace*{-0.8\tabcolsep}
S\cap T & \txtnach{\:\:\iota\:\:} 
                 & (S\cap T) \kreuz \rnkl H H   & \txtnach{\loktriv_T\loktriv_S^{-1}} 
                 & (S\cap T)\kreuz\rnkl{H}{N(H)}&\txtnach{\:\pr_2\:}&\rnkl{H}{N(H)}.
                                                       \hspace*{-0.8\tabcolsep}\\
   x    & \auf     &       (x    ,      H)        & \auf                      
                   & (x , H a_{\vartheta_x}^{-1}) & \auf        & \uebergabb^S_T(x)
\end{array}$%

}
\qed
\eunum
\epf

\section{Codimension of Nongeneric Strata}
\label{app:kodim_strata_mult}
Let $\LG$ be a compact Lie group acting smoothly on a manifold $M$.
By $(x_1,\ldots,x_k) \circ g := (x_1 \circ g, \ldots, x_k \circ g)$
for every $k\in\N_+$ we define 
an (again smooth) action of $\LG$ on $M^k$.
Obviously, the types of this action are again conjugacy classes
of closed subgroups of $\LG$.
Now we have 
\bprop
\label{prop:dim_strata_mult}
Let $\LG$ and $M$ be as above, $\LH$ be a closed subgroup of $\LG$ 
and $k\in\N_+$ be arbitrary.
Then we have
\zgl{\max_{[\LK]<[\LH]} \dim (M^k)_{\LK} 
     \leq k \max_{[\LK]<[\LH]} \dim M_{\LK}.}
\eprop
\bpf
Let $\vec x:=(x_1,\ldots,x_k) \in (M^k)_{\LK}$ for some $[\LK] < [\LH]$.
For some appropriate $g\in \LG$ we have
$g^{-1} \LK g = \Stab{\vec x} = \bigcap_i \Stab{x_i} \teilmenge \Stab{x_i}$
for all $i$.
Thus, $[\Stab{x_i}] \leq [\LK] < [\LH]$,
hence $x_i \in \bigcup_{[\LK_i]\leq [\LK]} M_{\LK_i}$.

Consequently,
\zgl{(M^k)_{\LK} \teilmenge 
         \dirprod_i \Bigl(\bigcup_{[\LK_i]\leq[\LK]} M_{\LK_i}\Bigr) 
       = \bigcup_{[\LK_i]\leq[\LK]} M_{\LK_1} \kreuz \cdots \kreuz M_{\LK_k}.}
  
Since there are only finitely many orbit types on $M^k$ \cite{EMS20}, we get
\bgl
\dim (M^k)_{\LK} 
 & \leq & \max_{[\LK_i]\leq[\LK]} 
                  \dim M_{\LK_1} \kreuz \cdots \kreuz M_{\LK_k} \\
 &  =   & k \: \max_{[\LK']\leq[\LK]} \dim M_{\LK'} \\
 & \leq & k \: \max_{[\LK']<[\LH]} \dim M_{\LK'}.
\egl
In particular, we have
\zgl{\max_{[\LK]<[\LH]} \dim (M^k)_{\LK} 
     \leq k \max_{[\LK]<[\LH]} \dim M_{\LK}.}
\qed
\epf
\bcorr
\label{corr:kodim_strata_mult}
We have
\zgl{\min_{[\LK]<[\LH]} \codim_{M^k} (M^k)_{\LK} 
     \geq k \min_{[\LK]<[\LH]} \codim_{M} M_{\LK}.}
\ecorr

\section{Proof of Proposition \ref{prop:ex(weak_hyph=typ)}}
\label{app:proof(prop:ex(weak_hyph=typ))}
First we discuss properties of so-called fundamental systems of
connected hyphs, i.e., certain free generating systems for the group of 
based paths spanned by that hyph.
For that purpose, we consider hyphs as abstract graphs. 
This makes graph-theoretical concepts applicable, 
like the notion of maximal trees.
\bdf
\bunum
\item
Let $\hyph = \{e_1,\ldots,e_\Hyph\}$ 
be a connected hyph with $m\in\Ver(\hyph)$
and $\hyph'$ be a maximal tree in $\hyph$.
Moreover, let $\phi : [1,n] \nach [1,\Hyph]$ be an injective
function, such that
$\hyph' = \hyph\setminus\{e_{\phi(1)},\ldots,e_{\phi(n)}\}$.
(This intricate definition is necessary, because 
$\hyph'$ is to be a hyph again and the hyph property
requires a certain ordering of the edges of a hyph.)

$\ga\teilmenge\hg$ is called 
\df{weak $\hyph'$-fundamental system for $\hyph$} iff
for every $i=1,\ldots,n$ the path $\alpha_i$ can be expressed as 
a product of three paths, whereas the first and the third path
are contained in 
$\KG{\{e_{\phi(1)},\ldots,e_{\phi(i-1)}\} \cup \hyph'}$ and
the second path equals the edge $e_{\phi(i)}$.

The edge $e_{\phi(i)}$ is called \df{free} edge of 
$\alpha_i$.
\item
$\ga\teilmenge\hg$ is called weak 
fundamental system iff there are a connected hyph $\hyph$ and a maximal 
tree $\hyph'$ in $\hyph$, such that $\ga$ is a weak 
$\hyph'$-fundamental system for $\hyph$.
\eunum
\edf
Here, $\KG\gc$ denotes the subgroupoid of $\Pf$ generated by 
$\gc \teilmenge \Pf$. Analogously, $\hg_\gb$ is defined.

We have obviously
\blem
\label{lem:ex(hyph_fundsyst)}
Every connected hyph possesses a weak fundamental system.
\elem
The basic properties of weak fundamental systems are summarized in 
\bprop
\label{prop:eig(hyph_fundsystem)}
Let $\hyph$ be a connected hyph with $m\in\Ver(\hyph)$,
and let $\ga$ be a weak fundamental system of $\hyph$. Then the
following holds:
\bnum{3}
\item
$\KG\ga$ is freely generated by $\ga$ and equals $\hg_\hyph$;
\item
$\pi_\ga{}_\ast\mu_0 = \mu_\Haar^{\elanz\ga}$;
\item
$\pi_\ga : \Ab \nach \LG^{\elanz\ga}$ is surjective.
\enum
Consequently, every weak fundamental system of a connected hyph is a weak hyph.
\eprop
\bpf
By assumption there is a maximal tree $\hyph'$ in 
$\hyph = \{e_1,\ldots,e_\Hyph\}$, such that 
$\ga$ is a $\hyph'$-fundamental system of $\hyph$. 
In order to avoid more complicated expressions,
we simply assume $\hyph' = \{e_{n+1},\ldots,e_\Hyph\}$.
\bnum{3}
\item
\bunum
\item
Obviously $\KG\ga\teilmenge\hg_\hyph$.
\item
We show $\hg_\hyph\teilmenge\KG\ga$ 
inductively w.r.t.\ $n = \elanz\ga = \elanz\hyph - \elanz\hyph'$.

For $n = 0$ the statement is trivial.
Now, let $n > 0$ and 
$\gamma\in\hg_\hyph$, i.e.\
$\gamma = \gamma_1 e^{\eta_1}_n \gamma_2 e^{\eta_2}_n \cdots 
                   e^{\eta_q}_n \gamma_{q+1}$
with $\gamma_j \in \KG{\widetilde\hyph}$ and $\eta_j = \pm 1$ 
for all $j$. Here $\widetilde\hyph := \hyph\setminus\{e_n\}$.
Let $\delta_{+1}$ and $\delta_{-1}$ be those paths in
$\KG{\widetilde\hyph}$ that fulfill
$\alpha_n = \delta_{-1} e_n \delta_{+1}$.
Hence,
\zglnum{\gamma = \klammerunten{\gamma_1 \delta_{-\eta_1}^{-\eta_1}}%
                           {\in\hg_{\widetilde\hyph}}
              \alpha^{\eta_1}_n 
              \klammerunten{\delta_{\eta_1}^{-\eta_1} \gamma_2 
                            \delta_{-\eta_2}^{-\eta_2}}%
                           {\in\hg_{\widetilde\hyph}}
              \alpha^{\eta_2}_n \:
              \delta_{\eta_2}^{-\eta_2}
              \cdots
              \delta_{-\eta_q}^{-\eta_q} \:
              \alpha^{\eta_q}_n 
              \klammerunten{\delta_{\eta_q}^{-\eta_q}\gamma_{q+1}}%
                           {\in\hg_{\widetilde\hyph}}
              .}{gl:fundsyst}
By $e_n\nichtin\hyph'$, also $\hyph'$ is a maximal tree
for $\widetilde\hyph$ and, in particular, $\widetilde\hyph$ is connected.
Now, $\ga\setminus\{\alpha_n\}$ is a weak
$\hyph'$-fundamental system for $\widetilde\hyph$.
Due to
$\elanz\widetilde\hyph - \elanz\hyph' = \elanz\hyph - \elanz\hyph' - 1 = n - 1$
we have by the induction hypothesis
$\hg_{\widetilde\hyph} = \KG{\ga\setminus\{\alpha_n\}}$.
Equation \eqref{gl:fundsyst} implies
$\gamma \in \KG{\ga}$.
\item
Consequently, $\ga$ is a generating system of the free group $\hg_{\hyph}$,
whose rank equals $\elanz\hyph - \elanz\hyph'$, hence $\elanz\ga$.
Thus, $\ga$ is free. \cite{Kurosch1}
\eunum
\item
We have 
\fktdefabgesetzt{\pi^\hyph_\ga}{\LG^\Hyph}{\LG^n,}
                               {(g_1,\ldots,g_\Hyph)}
                               {(\cdots g_1 \cdots, \ldots, \cdots g_n \cdots)}
where the $\cdots$ in $\cdots g_i \cdots$ denote a product of some
$g_j$ or $g_j^{-1}$ with $j > n$ or $j < i$.
Now, we get for all $f \in C(\LG^n)$
using successively the structure of 
$\cdots g_i \cdots$ and the translation invariance and the normalization 
of the Haar measure
\bgl[2ex]
\zurueck 
 &   & \int_{\LG^\Hyph} (\pi^\hyph_\ga)^\ast f 
                           \: \dd\mu_\Haar^\Hyph \s
\zurueck 
 & = & \int_{\LG^\Hyph} f(\pi^\hyph_\ga(g_1,\ldots,g_\Hyph))
                           \: \dd\mu_{\Haar,1} \cdots \dd\mu_{\Haar,\Hyph} \s
\zurueck 
 & = & \int_{\LG^\Hyph} f(\cdots g_1 \cdots,\ldots,
                               \cdots g_{n-1} \cdots,\cdots g_n \cdots)
                           \: \dd\mu_{\Haar,1} \cdots \dd\mu_{\Haar,\Hyph} \s
\zurueck 
 & = & \int_{\LG^\Hyph} f(\cdots g_1 \cdots,\ldots,\cdots g_{n-1} \cdots,g_n)
                           \: \dd\mu_{\Haar,1} \cdots \dd\mu_{\Haar,\Hyph} \s
\zurueck 
 & \vdots  & \s
\zurueck 
 & = & \int_{\LG^n} f(g_1,\ldots,g_n)
                           \: \dd\mu_{\Haar,1} \cdots \dd\mu_{\Haar,n} \s
\zurueck 
 & = & \int_{\LG^n} f \: \dd\mu_\Haar^n.
\egl
Since $\pi_\hyph$ projects the induced Haar measure onto the Haar measure
\cite{paper3}, we have
$\pi_\ga{}_\ast \mu_0 = (\pi^\hyph_\ga)_\ast (\pi_\hyph{}_\ast \mu_0)
                      = (\pi^\hyph_\ga)_\ast \mu_\Haar^\Hyph
                      = \mu_\Haar^n$.
\item
Let $(g_1,\ldots,g_n) \in \LG^n$.
By assumption we can express every 
$e_i$, $i\leq n$, as a product of $\alpha_i$ with appropriate paths in
$\KG{\{e_{1},\ldots,e_{i-1}\} \cup \hyph'}$.
Now, we choose $\delta_{i,\pm}\in\KG{\hyph'}$, such that
$\delta_{i,-} e_i \delta_{i,+}$ is a path in $\hg$.
But, then $\delta_{i,-} e_i \delta_{i,+}$ equals a product of a closed path in
$\KG{\{e_{1},\ldots,e_{i-1}\} \cup \hyph'}$, the path $\alpha_i$ and
a further closed path 
$\KG{\{e_{1},\ldots,e_{i-1}\} \cup \hyph'}$.
By the first step $\hg_{\{e_{1},\ldots,e_{i-1}\} \cup \hyph'} 
 = \KG{\{\alpha_1,\ldots,\alpha_{i-1}\}}$.
Hence, 
$\delta_{i,-} e_i \delta_{i,+} 
  = \prod_{k_i}^{K_i} \alpha_{k(k_i,i)}^{\pm 1}
    \alpha_i  
    \prod_{l_i}^{L_i} \alpha_{l(l_i,i)}^{\pm 1}$
for some functions $k$ and $l$, that fulfill always
$k(k_i,i) < i$ and $l(l_i,i) < i$.

Since $\pi_\hyph$ is surjective for every hyph \cite{paper3},
there is an $\qa\in\Ab$, such that 
$h_\qa(e_i) = e_\LG$ for all $i > n$ and
$h_\qa(e_i) = h_\qa(\delta_{i,-})^{-1} \:
              \bigl(\prod_{k_i}^{K_i} g_{k(k_i,i)}^{\pm 1}\bigr)
              \: g_i \:
              \bigl(\prod_{l_i}^{L_i} g_{l(l_i,i)}^{\pm 1}\bigr)
              \: h_\qa(\delta_{i,+})^{-1}$
for all $i\leq n$ (defined inductively, since $\delta_{i,\pm}$
may run through $e_{i'}$ with $i'<i$ or $i'>n$).
By construction we have 
$h_\qa(\alpha_i) = g_i$ for all $i=1, \ldots, n$.
\qed
\enum
\epf
Finally, we have
\bpf[Proposition \ref{prop:ex(weak_hyph=typ)}]
First we choose a $\gb\teilmenge\hg$ with
$Z(h_\qa(\gb)) = Z(\holgr_\qa)$.
Since $\gb$ is obviously connected,
there is a connected hyph $\hyph$, such that
every path in $\gb$ can be expressed by paths in $\hyph$.
(The existence of a hyph is proven in \cite{paper3}, the connectedness 
is a simple consequence \cite{diss}.)
Lemma \ref{lem:ex(hyph_fundsyst)} guarantees the existence of a 
weak fundamental system
$\ga\teilmenge\hg$ for $\hyph$. By Proposition \ref{prop:eig(hyph_fundsystem)},
$\ga$ is a weak hyph. 
Since every $\beta\in\hg$ is a closed path in $\hyph$ by construction, 
it can be written as a product of paths in $\ga$ (and their inverses).
Hence, $h_\qa(\beta)$ is a product of $h_\qa(\alpha_i)$ and their inverses. 

Now, let $g\in Z(h_\qa(\ga))$. Then $g$ commutes with all 
finite products of $h_\qa(\alpha_i)$ and their inverses as well;
thus, in particular, $g\in Z(h_\qa(\beta))$.
This implies 
$Z(\holgr_\qa) \teilmenge Z(h_\qa(\ga)) \teilmenge 
 \bigcap_{\beta\in\gb} Z(h_\qa(\beta)) = Z(h_\qa(\gb)) = Z(\holgr_\qa)$,
hence $Z(h_\qa(\ga)) = Z(\holgr_\qa)$.
\qed
\epf


\section{Homeomorphy Criterion}
\label{app:topology}
\bprop
\label{prop:stetinversschwach}
Let $X$ be a topological space, $Y$ be a Hausdorff space and 
$f:X\nach Y$ be a bijective and continuous mapping.

If every point in $Y$ has a compact neighbourhood whose preimage 
is compact again, then $f$ is a homeomorphism.
\eprop
\bpf
Let $x\in X$. By assumption there is a compact 
neighbourhood $W\teilmenge Y$ of $f(x)$ such that
$V:=f^{-1}(W)$ is compact again. 
By continuity of $f$, $V$ is a neighbourhood of $x$.
Hence, $f\einschr V:V\nach W$ is a bijective and continuous
mapping of a compact space into a Hausdorff space, hence a
homeomorphism.

Now we show that $f^{-1}$ is continuous in $f(x)$.
For this, let $U$ be an open neighbourhood of $x$ in $X$ and $W_0\teilmenge W$
be an open neighbourhood of $f(x)$ in $Y$. If one sets $V_0:=f^{-1}(W_0)$,
then $U\cap V_0$ is an open neighbourhood of $x$ in $X$, hence 
in $V$ as well.
Consequently, $f\einschr V(U\cap V_0) = f(U\cap V_0)$ is open w.r.t.\
$f(V) = W$. By $f(V_0) = W_0$ and the bijectivity of
$f$ we have $f(U) \cap W_0 = f(U) \cap f(V_0) = f(U\cap V_0) \teilmenge W_0$.
But, $f(U)\cap W_0$ is open in $W$, hence in $W_0$ as well and thus 
(by the openness of $W_0$ in $Y$) in $Y$, too.
Hence, $f(U)\obermenge f(U)\cap W_0 \ni f(x)$ is a neighbourhood
of $f(x)$ in $Y$. 
Since this is true for all open neighbourhoods $U$ of $x$, $f^{-1}$ 
is continuous.
\qed
\epf


\end{document}